\documentclass[onecolumn,nofootinbib,superscriptedaddress,notitlepage,amsmath,amssymb,fleqn,prd,preprintnumbers]{revtex4-1}
\usepackage{xcolor,braket,graphicx,xspace,todonotes}

\newcommand{\eps}{\varepsilon}
\newcommand{\Alaric}{{\sc Alaric}\xspace}
\newcommand{\Dire}{{\sc Dire}\xspace}
\newcommand{\Sherpa}{{\sc Sherpa}\xspace}
\usepackage{hyperref}

\hypersetup{hidelinks,
  pdfauthor={Benoit Assi,Stefan Hoeche},
  pdftitle={A new approach to QCD final-state evolution in processes with massive partons},
  pdfkeywords={Parton Shower, QCD Resummation}
}

\begin{document}
\preprint{FERMILAB-PUB-23-336-T, MCNET-23-09}
\title{A new approach to QCD final-state evolution in processes with massive partons}
\author{Beno{\^i}t Assi}
\affiliation{Fermi National Accelerator Laboratory, Batavia, IL, 60510}
\author{Stefan H{\"o}che}
\affiliation{Fermi National Accelerator Laboratory, Batavia, IL, 60510}

\begin{abstract}
    We present an algorithm for massive parton evolution which is based on the
    differentially accurate simulation of soft-gluon radiation by means of a
    non-trivial azimuthal angle dependence of the splitting functions. 
    The kinematics mapping is chosen such as to to reflect the symmetry 
    of the final state in soft-gluon radiation and collinear splitting processes.
    We compute the counterterms needed for a fully differential NLO matching
    and discuss the analytic structure of the parton shower in the NLL limit.
    We implement the new algorithm in the numerical code \Alaric
    and present a first comparison to experimental data.
\end{abstract}

\maketitle

\section{Introduction}
The production and evolution of massive partons are an important aspect of collider physics, 
and they play a particularly prominent role at the Large Hadron Collider at CERN. Key measurements
and searches, such as $t\bar{t}H$ and double Higgs boson production, involve final states with many
$b$-jets. The success of the LHC physics program therefore depends crucially on the modeling
of heavy quark processes in the Monte-Carlo event generators used to link theory and experiment.
With the high-luminosity phase of the LHC approaching fast, it is important to increase
the precision of these tools in simulations involving massive partons.

Heavy quark and heavy-quark associated processes have been investigated in great detail, 
both from the perspective of fixed-order perturbative QCD and using resummation, 
see for example~\cite{Mangano:1991jk,Bonciani:1998vc,Cacciari:2003uh,Cacciari:2012ny}.
Various proposals were made for the fully differential simulation in the context of particle-level
Monte Carlo event generators~\cite{Norrbin:2000uu,Gieseke:2003rz,Schumann:2007mg,Gehrmann-DeRidder:2011gkt}.
Recently, a new scheme was devised for including the evolution of massive quarks in the initial state
of hadron-hadron and lepton-hadron collisions~\cite{Krauss:2017wmx}.
In this manuscript, we will introduce an algorithm for the final-state evolution and matching in
heavy-quark processes, inspired by the recently proposed parton-shower model \Alaric~\cite{Herren:2022jej}.
The soft components of the splitting functions are derived from the massive
eikonal and are matched to the quasi-collinear limit using a partial fractioning technique.
In contrast to the method of~\cite{Catani:2002hc,Schumann:2007mg}, we partial fraction the complete
soft eikonal, leading to strictly positive splitting functions and thus keeping the numerical 
efficiency of the Monte-Carlo algorithm at a maximum. We also propose to use a kinematic mapping
for the collinear splitting of gluons into quarks that treats the outgoing particles democratically.
This algorithm can be extended to any purely collinear splitting (i.e., \ after subtracting any soft
enhanced part of the splitting functions) while retaining the NLL precision of the evolution.
Our algorithm does not account for the effect of spin correlations.

Multi-jet merging and matching of parton-shower simulations to NLO calculations in the context
of heavy-quark production were discussed, for example, in~\cite{Mangano:2001xp,Frixione:2003ei,
  Frixione:2007nw,Hoeche:2013mua}. The NLO matching is typically fairly involved, because
of the complex structure and partly ambiguous definition of the infrared counterterms.
In this publication, we compute the integrated counterterms for our new parton-shower model,
making use of recent results for angular integrals in dimensional regularization~\cite{Lyubovitskij:2021ges}.
This calculation provides the remaining counterterms needed for the matching of the \Alaric
parton-shower model at NLO QCD. We will discuss the extension to initial-state radiation
in a future publication.

This paper is structured as follows: Section~\ref{sec:coherence} briefly reviews the construction
of the \Alaric parton-shower model and generalizes the discussion to massive
particles. Section~\ref{sec:kinematics} introduces the different kinematics mappings.
Section~\ref{sec:ps_factorization} discusses the general form of the phase-space factorization
and provides explicit results for processes with soft radiation and collinear splitting.
The computation of integrated infrared counterterms is presented in Sec.~\ref{sec:nlo_subtraction}.
Section~\ref{sec:nll_proof} discusses the impact of the kinematics mapping on sub-leading logarithms,
and Sec.~\ref{sec:comparison} provides first numerical predictions for $e^+e^-\to$hadrons.
Section~\ref{sec:conclusions} contains an outlook.

\section{Soft-collinear matching}
\label{sec:coherence}
We start the discussion by revisiting the singularity structure of $n$-parton
QCD amplitudes in the infrared limits.
If two partons, $i$ and $j$, become quasi-collinear, the squared amplitude factorizes as
\begin{equation}
    _{n}\langle1,\ldots,n|1,\ldots,n\rangle_{n}=
    \sum_{\lambda,\lambda'=\pm}
    \, _{n-1}{\Big<}1,\ldots,i\!\!\backslash(ij),\ldots,j\!\!\!\backslash,\ldots,n\Big|
    \frac{8\pi\alpha_s\,P^{\lambda\lambda'}_{(ij)i}(z)}{(p_i+p_j)^2-m_{ij}^2}
    \Big|1,\ldots,i\!\!\backslash(ij),\ldots,j\!\!\!\backslash,\ldots,n\Big>_{n-1}\;,
\end{equation}
where the notation $i\!\!\backslash$ indicates that parton $i$ is removed 
from the original amplitude, and where $(ij)$ is the progenitor of partons $i$ and $j$.
The functions $P^{\lambda\lambda'}_{ab}(z)$ are the spin-dependent, massive DGLAP splitting functions,
which depend on the momentum fraction $z$ of parton $i$ with respect to the mother parton, $(ij)$,
and on the helicities $\lambda$~\cite{Dokshitzer:1977sg,Gribov:1972ri,Lipatov:1974qm,Altarelli:1977zs,Catani:2000ef,Catani:2002hc}.
For the remainder of this work, we will use only the spin-averaged splitting functions.

In the limit that gluon $j$ becomes soft, the squared amplitude factorizes as~\cite{Bassetto:1984ik}
\begin{equation}
    _{n}\langle1,\ldots,n|1,\ldots,n\rangle_{n}=-8\pi\alpha_s\sum_{i,k\neq i}
    \,_{n-1}\big<1,\ldots,j\!\!\!\backslash,\ldots,n\big|{\bf T}_i{\bf T}_k\,w_{ik,j}
    \big|1,\ldots,j\!\!\!\backslash,\ldots,n\big>_{n-1}\;,
\end{equation}
where ${\bf T}_i$ and ${\bf T}_k$ are the color insertion operators defined in~\cite{Bassetto:1984ik,Catani:1996vz}.
In the remainder of this section, we will focus on the eikonal factor, $w_{ik,j}$, for massive partons
and how it can be re-written in a suitable form to match the spin-averaged splitting functions
in the soft-collinear limit. The eikonal factor is given by
\begin{equation}\label{eq:massive_eikonal}
  \begin{split}
    w_{ik,j}=&\;\frac{p_ip_k}{(p_ip_j)(p_jp_k)}-\frac{p_i^2/2}{(p_ip_j)^2}-\frac{p_k^2/2}{(p_kp_j)^2} \\
  \end{split}
\end{equation}
Following Refs.~\cite{Webber:1986mc,Herren:2022jej}, we split Eq.~\eqref{eq:massive_eikonal}
into an angular radiator function, and the gluon energy
\begin{equation}\label{eq:massive_eikonal_split}
    w_{ik,j}=\frac{W_{ik,j}}{E_j^2}\;,
    \quad\text{where}\quad
    W_{ik,j}=\frac{1-v_iv_k\cos{\theta_{ik}}}{(1-v_i\cos{\theta_{ij}})(1-v_k\cos{\theta_{jk}})}
        -\frac{(1-v_i^2)/2}{(1-v_i\cos{\theta_{ij}})^2}
        -\frac{(1-v_k^2)/2}{(1-v_k\cos{\theta_{jk}})^2}\;.
\end{equation}
The parton velocity is defined as $v=\sqrt{1-m^2/E^2}$ and is frame dependent.
When we label velocities by particle indices in the following, it is implicit that 
they are computed in the frame of a time-like momentum $n^\mu$. In this reference frame
they reduce to the relative velocity $v_i=v_{p_in}$, where
\begin{equation}\label{eq:relative_velocity}
  v_{pq}=\sqrt{1-\frac{p^2q^2}{(pq)^2}}\;.
\end{equation}
For convenience, we also introduce the velocity four-vector
\begin{equation}\label{eq:velocity_vectors}
    l_{p,q}^\mu=\sqrt{q^2}\,\frac{p^\mu}{pq}\;.
\end{equation}
When this vector is labeled by a single particle index only, it again refers to the four-velocity
of that particle in the frame of $n^\mu$, for example $l_i^\mu=l_{p_i,n}^\mu$.
When partial fractioning Eq.~\eqref{eq:massive_eikonal_split}, we aim at a positive definite result
in order to maintain the interpretation of a probability density and, with it, the high efficiency
of the unweighted event generation in a parton shower. Following the approach of 
Ref.~\cite{Herren:2022jej}, we obtain
\begin{equation}
    \label{eq:partfrac_soft_matching}
    W_{ik,j}=\bar{W}_{ik,j}^i+\bar{W}_{ki,j}^k\;,
    \qquad\text{where}\qquad
    \bar{W}_{ik,j}^i=\frac{1-v_k\cos\theta_{jk}}{
        2-v_i\cos\theta_{ij}-v_k\cos\theta_{jk}}\,
      W_{ik,j}\;.
\end{equation}
The partial fractioned angular radiator function can be written in a more convenient form
using the velocity four-vectors. We find an expression that makes the matching to the
$ij$-collinear sector manifest
\begin{equation}\label{eq:pf_angular_radiator}
    \bar{W}_{ik,j}^i=\frac{1}{2l_il_j}\left(\frac{l_{ik}^2}{l_{ik}l_j}
        -\frac{l_i^2}{l_il_j}-\frac{l_k^2}{l_kl_j}\right)\;,
    \qquad\text{where}\qquad
    l_{ik}^\mu=l_i^\mu+l_k^\mu\;.
\end{equation}
In the quasi-collinear limit for partons $i$ and $j$, we can write 
the eikonal factor in Eq.~\eqref{eq:massive_eikonal} as~\cite{Catani:2000ef,Catani:2002hc}
\begin{equation}
  w_{ik,j}\underset{m_i\varpropto p_ip_j}{\overset{i||j}{\longrightarrow}}
  w_{ik,j}^{\rm(coll)}(z)=\frac{1}{2p_ip_j}\left(\frac{2z}{1-z}-\frac{m_i^2}{p_ip_j}\right)\;,
  \qquad\text{where}\qquad
  z\underset{m_i\varpropto p_ip_j}{\overset{i||j}{\longrightarrow}}\frac{E_i}{E_i+E_j}\;.
\end{equation}
This can be identified with the leading term (in $1-z$) of the
DGLAP splitting functions $P_{aa}(z,\eps)$, where
\footnote{
Note that in contrast to standard DGLAP notation, we separate the gluon
splitting function into two parts, associated with the soft singularities
at $z\to 0$ and $z\to1$.}
\begin{equation}\label{eq:dglap_splittings}
    \begin{split}
        P_{qq}(z,\eps)&=C_F\left(\frac{2z}{1-z}-\frac{m_i^2}{p_ip_j}+(1-\eps)(1-z)\right)\;,\\
        P_{gg}(z,\eps)&=C_A\left(\frac{2z}{1-z}+z(1-z)\right)\;,\\
        P_{gq}(z,\eps)&=T_R\left(1-\frac{2z(1-z)}{1-\eps}\right)\;.
    \end{split}
\end{equation}
To match the soft to the collinear splitting functions,
we therefore replace
\begin{equation}\label{eq:matching_soft_coll}
    \frac{P_{(ij)i}(z,\eps)}{(p_i+p_j)^2-m_{ij}^2}\to
    \frac{P_{(ij)i}(z,\eps)}{(p_i+p_j)^2-m_{ij}^2}+\delta_{(ij)i}\,{\bf T}_{i}^2
    \Bigg[\frac{\bar{W}_{ik,j}^i}{E_j^2}-w_{ik,j}^{\rm(coll)}(z)\Bigg]\;,
\end{equation}
where the two contributions to the gluon splitting function are treated 
as two different radiators~\cite{Hoche:2015sya}. As in the massless case,
this substitution introduces a dependence on a color spectator, $k$, 
whose momentum defines a direction independent of the direction of the 
collinear splitting~\cite{Herren:2022jej}. In general, this implies that the
splitting functions which were formerly dependent only on a momentum fraction 
along this direction, now acquire a dependence on the remaining two phase-space 
variables of the new parton. It is convenient to define the purely collinear
remainder functions
\begin{equation}\label{eq:coll_remainder}
    \begin{split}
        C_{qq}(z,\eps)&=C_F(1-\eps)\left(1-z\right)\;,\\
        C_{gg}(z,\eps)&=C_A\,z(1-z)\;,\\
        C_{gq}(z,\eps)&=T_R\left(1-\frac{2z(1-z)}{1-\eps}\right)\;.
    \end{split}
\end{equation}
They can be implemented in the parton shower using collinear kinematics,
while maintaining the overall NLL precision of the simulation. This will be
discussed further in Sec.~\ref{sec:nll_proof}.

\section{Momentum mapping}
\label{sec:kinematics}
Every parton shower algorithm requires a method to map the momenta of the
Born process to a kinematical configuration after additional parton emission.
This mapping is linked to the factorization of the differential phase-space
element for a multi-parton configuration. Collinear safety is a basic 
requirement for every momentum mapping. In addition, a mapping is NLL safe
if it preserves the topological features of radiation simulated
previously~\cite{Dasgupta:2018nvj,Dasgupta:2020fwr}.
We will make this statement more precise in Sec.~\ref{sec:nll_proof}.

This section provides a generic momentum mapping for massive partons
that is both collinear and NLL-safe. It is constructed for both initial-
and final-state radiation, inspired by the identified particle 
dipole subtraction algorithm of~\cite{Catani:1996vz}. In the massless limit,
we reproduce this algorithm and thus agree with the existing parton-shower model
\Alaric~\cite{Herren:2022jej}.

\subsection{Radiation kinematics}
\label{sec:radiation_kinematics}
\begin{figure}[t]
\includegraphics[width=\textwidth]{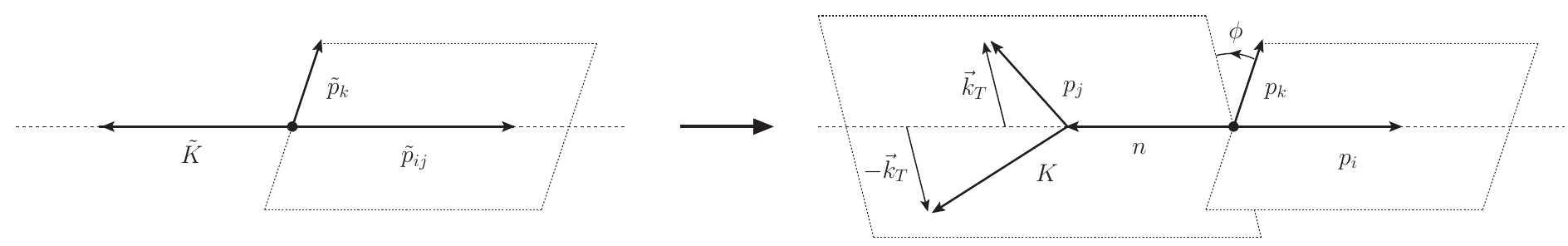}
\caption{Sketch of the radiation kinematics in final-state evolution.
See the main text for details. Note that $p_k$ is unaltered by the mapping
and only acts as a reference for the azimuthal angle $\phi$
(cf.\ Eqs.~\eqref{eq:def_n_perp} and~\eqref{eq:def_l_perp}).
\label{fig:ff_rad_kinematics}}
\end{figure}

We will first describe the kinematics mapping needed for soft evolution.
This is sketched in Fig.~\ref{fig:ff_rad_kinematics}.
The momenta $\tilde{K}$ and $\tilde{p}_{ij}$ are mapped to the momenta
$K$, $p_i$ and $p_j$, while $\tilde{p}_k$ acts as a spectator that defines
the azimuthal direction. We assume that any of the particles $i$, $j$
and $k$ can be massive. The momentum $\tilde{K}$ can be chosen in a suitable
way to reflect the dynamics of the process~\cite{Herren:2022jej} and absorbs
the recoil in the splitting. We define the variables
\begin{equation}
  \mu_{ij}^2=\frac{\tilde{p}_{ij}^2}{2\tilde{p}_{ij}\tilde{K}}\;,\qquad
  \bar{\mu}_{ij}^2=\frac{2\mu_{ij}^2}{1+v_{\tilde{p}_{ij}\tilde{K}}}\;,\qquad
  \kappa=\frac{\tilde{K}^2}{2\tilde{p}_{ij}\tilde{K}}\;,\qquad
  \bar{\kappa}=\frac{2\kappa}{1+v_{\tilde{p}_{ij}\tilde{K}}}\;,
\end{equation}
and the analogous final-state variables
$\mu_{i/j}^2=m_{i/j}^2/(2\tilde{p}_{ij}\tilde{K})$ and
$\bar{\mu}_{i/j}^2=2\mu_{i/j}^2/(1+v_{\tilde{p}_{ij}\tilde{K}})$.\\
The scalar invariants after the splitting are defined in terms of 
the energy fraction, $z$ \cite{Catani:1996vz}
\begin{equation}\label{eq:def_z_n_rad}
  2p_in=z\,2\tilde{p}_{ij}\tilde{K}\;,\qquad
  n^2=\big(1-z+\kappa+\mu_{ij}^2-\mu_i^2\big)\,2\tilde{p}_{ij}\tilde{K}\;.
\end{equation}
The momentum of the radiator after the emission is
\begin{equation}\label{eq:def_pi_rad}
  p_i^\mu=\bar{z}\,\tilde{p}_{ij}^\mu+\frac{\mu_i^2-\bar{z}^2\mu_{ij}^2}{
  \bar{z}\,v_{\tilde{p}_{ij}\tilde{K}}}\left(\tilde{K}^\mu-\bar{\kappa}\,\tilde{p}_{ij}^\mu\right)\;,
\end{equation}
where
\begin{equation}
  \bar{z}=\frac{z+2\mu_i^2}{1+v_{\tilde{p}_{ij}\tilde{K}}+2\mu_{ij}^2}
  +\sqrt{\bigg(\frac{z+2\mu_i^2}{1+v_{\tilde{p}_{ij}\tilde{K}}+2\mu_{ij}^2}\bigg)^2
  -\frac{2\mu_i^2(1+\bar{\kappa})}{1+v_{\tilde{p}_{ij}\tilde{K}}+2\mu_{ij}^2}}\;.
\end{equation}
We define the variable
\begin{equation}
    \bar{v}=
    \frac{\displaystyle \frac{2\,v\,z}{1+v_{\tilde{p}_{ij}\tilde{K}}}
    -\frac{\bar{\mu}_i^2}{\bar{z}}\bigg(1-\bar{z}+\bar{\kappa}-\frac{\bar{\kappa}-\bar{\mu}_j^2}{\zeta}\bigg)}{
    \displaystyle \bar{z}\,-\frac{\bar{\mu}_i^2}{\bar{z}}\frac{1-\bar{z}+\bar{\kappa}}{\zeta}}\;,
    \qquad\text{where}\qquad
    \zeta=\frac{2}{1+v_{\tilde{p}_{ij}\tilde{K}}}\,
    \frac{1-z+\kappa+\mu_{ij}^2-\mu_i^2}{1-\bar{z}+\bar{\kappa}}\;,
\end{equation}
and where $v=(p_ip_j)/(p_in)$ is defined as in the massless case~\cite{Catani:1996vz,Herren:2022jej}. 
In terms of these quantities, the gluon momentum is given by
\begin{equation}\label{eq:def_pj_rad}
  p_j^\mu=\frac{1-z+\mu_{ij}^2-\mu_i^2+\mu_j^2}{v_{\tilde{p}_{ij}\tilde{K}}\zeta}\left(\tilde{p}_{ij}^\mu-\bar{\mu}_{ij}^2\tilde{K}^\mu\right)
  +\bar{v}\,\frac{1+v_{\tilde{p}_{ij}\tilde{K}}}{2v_{\tilde{p}_{ij}\tilde{K}}}\left[\left(\tilde{K}^\mu-\bar{\kappa}\tilde{p}_{ij}^\mu\right)
  -\frac{1-\bar{z}+\bar{\kappa}}{\zeta}\left(\tilde{p}_{ij}^\mu-\bar{\mu}_{ij}^2\tilde{K}^\mu\right)\right]
  +k_\perp^\mu\;.
\end{equation}
where
\begin{equation}
    {\rm k}_\perp^2=\tilde{p}_{ij}\tilde{K}(1+v_{\tilde{p}_{ij}\tilde{K}})\,
    \bar{v}\left[\left(1-\frac{\bar{v}}{\zeta}\right)(1-\bar{z})
    -\frac{1-\zeta+\bar{v}}{\zeta}\,\bar{\kappa}+\frac{\bar{\mu}_j^2}{\zeta}\,\right]-m_j^2\;.
\end{equation}
In order to determine a reference direction for the azimuthal angle
$\phi=\arctan({\rm k}_y/{\rm k}_x)$, we note that the soft radiation pattern
must be correctly generated. To achieve this, we compose the transverse momentum as
\begin{equation}\label{eq:likt}
  k_\perp^\mu={\rm k}_\perp\left(\cos\phi\, \frac{n_\perp^\mu}{|n_\perp|}
  +\sin\phi\, \frac{l_\perp^{\,\mu}}{|l_\perp|}\right)\;,
\end{equation}
where the reference axes $n_\perp$ and $l_\perp$ are given 
by the transverse projections\footnote{In kinematical configurations
  where $p_k^{\,\mu}$ is a linear combination of $p_i^{\,\mu}$ and
  $\bar{n}^{\,\mu}$, $n_\perp$ in the definition of Eq.~\eqref{eq:likt} vanishes. 
  It can then be computed using $n_\perp=\eps^{\mu j}_{\;\;\;\nu\rho}\, p_i^{\,\nu}\,\bar{n}^{\,\rho}$,
  where $j\in\{1,2,3\}$ may be any index that yields a nonzero result. 
  Note that in this case, the matrix element cannot depend on the azimuthal angle.}
\begin{equation}\label{eq:def_n_perp}
  n_\perp^\mu=p_k^\mu
  -\frac{\big(p_k(\tilde{K}-\bar{\kappa}\tilde{p}_{ij})\big)\,
  \big(\tilde{p}_{ij}^\mu-\bar{\mu}_{ij}^2\tilde{K}^\mu\big)}{
  2\tilde{p}_{ij}\tilde{K}\,v_{\tilde{p}_{ij}\tilde{K}}^2/(1+v_{\tilde{p}_{ij}\tilde{K}})}
  -\frac{\big(p_k(\tilde{p}_{ij}-\bar{\mu}_{ij}^2\tilde{K})\big)\,
  \big(\tilde{K}^\mu-\bar{\kappa}\tilde{p}_{ij}^\mu\big)}{
  2\tilde{p}_{ij}\tilde{K}\,v_{\tilde{p}_{ij}\tilde{K}}^2/(1+v_{\tilde{p}_{ij}\tilde{K}})}\;,
\end{equation}
and
\begin{equation}\label{eq:def_l_perp}
  l_\perp^{\,\mu}=\eps^\mu_{\;\nu\rho\sigma}\,
  \tilde{p}_{ij}^{\,\nu}\,
  \big(\tilde{K}^\rho-\bar{\kappa}\tilde{p}_{ij}^{\,\rho}\big)\,n_\perp^\sigma\;.
\end{equation}
Since the differential emission phase-space element is a Lorentz-invariant quantity,
the azimuthal angle $\phi$ is Lorentz invariant~\cite{Herren:2022jej}. This allows us
to write the emission phase-space in a frame-independent way. Upon construction 
of the momenta $p_i$, $p_j$ and $K$, the momenta $\{p_l\}$ which are used to define
$\tilde{K}$ are subjected to a Lorentz transformation,
\begin{equation}\label{eq:lorentz_transformation}
    p^{\mu}_l\rightarrow \Lambda^{\mu}_{\nu}(K,\tilde{K})p^{\nu}_l,
    \quad {\rm where} \quad
    \Lambda^{\mu}_{\nu}=g^{\mu}_{\nu}
    -2\frac{(\tilde{K}+K)^{\mu}(\tilde{K}+K)_{\nu}}{(\tilde{K}+K)^2}
    +2\frac{\tilde{K}^{\mu}K_{\nu}}{K^2}
\end{equation}

\subsection{Splitting kinematics}
\label{sec:splitting_kinematics}
\begin{figure}[t]
\includegraphics[width=\textwidth]{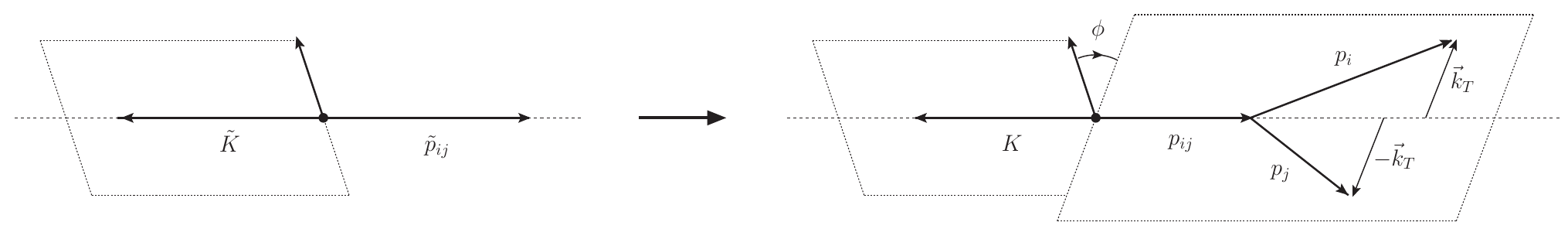}
\caption{Sketch of the splitting kinematics in final-state evolution.
See the main text for details. Note that the unlabeled momentum is unaltered by the mapping,
and only acts as a reference direction to define the azimuthal angle $\phi$.
\label{fig:ff_split_kinematics}}
\end{figure}
In this section, we describe the kinematics for the implementation of the purely collinear
components of the splitting functions in Eq.~\eqref{eq:coll_remainder}. 
This is sketched in Fig.~\ref{fig:ff_split_kinematics}.
We make use of some of the notation in~\cite{Catani:2002hc}, in particular
\begin{equation}\label{eq:def_y_z_cdst}
    y=\frac{p_ip_j}{p_ip_j+p_iK+p_jK}\;,
    \qquad
    z=\frac{p_iK}{p_iK+p_jK}\;,
\end{equation}
and we define the scaled masses
\begin{equation}
    \mu_i^2=\frac{m_i^2}{2\tilde{p}_{ij}\tilde{K}}\;,\qquad
    \mu_j^2=\frac{m_j^2}{2\tilde{p}_{ij}\tilde{K}}\;,\qquad
    \kappa=\frac{\tilde{K}^2}{2\tilde{p}_{ij}\tilde{K}}\;,\qquad
    \bar{\kappa}=\frac{2\kappa}{1+v_{\tilde{p}_{ij},\tilde{K}}}\;.
\end{equation}
We also introduce the following variables
\begin{equation}
    \alpha_{ij}=y+(1-y)(\mu_i^2+\mu_j^2)\;,\qquad
    \bar{\alpha}_{ij}=\frac{2\alpha_{ij}}{1+v_{\tilde{p}_{ij},\tilde{K}}}\;,\qquad
    \mu_{ij}^2=\frac{m_{ij}^2}{2\tilde{p}_{ij}\tilde{K}}\;,\qquad
    \bar{\mu}_{ij}^2=\frac{2\mu_{ij}^2}{1+v_{\tilde{p}_{ij},\tilde{K}}}\;.
\end{equation}
In terms of the additional variables
\begin{equation}
  \begin{split}
    z_{ij}=&\;\frac{1}{2\bar{\alpha}_{ij}}\Bigg(
    \frac{1+\bar{\alpha}_{ij}}{1+\bar{\kappa}}+\bar{\mu}_{ij}^2
    -\sqrt{\left(\frac{1-\bar{\alpha}_{ij}}{1+\bar{\kappa}}+\bar{\mu}_{ij}^2\right)^2
      -\frac{4\bar{\alpha}_{ij}\bar{\kappa}}{(1+\bar{\kappa})^2}}\;\Bigg)\;,\qquad
    \bar{z}_{ij}=\frac{2z_{ij}}{1+v_{\tilde{p}_{ij},\tilde{K}}}\;,\\
    \bar{z}=&\;\frac{\displaystyle z\big(1-\alpha_{ij}+\mu_{ij}^2\big)
    -\big(\alpha_{ij}+\mu_i^2-\mu_j^2\big)
    \frac{\bar{z}_{ij}\kappa}{1-\bar{z}_{ij}\alpha_{ij}+\bar{\mu}_{ij}^2}}{\displaystyle
    \frac{1-\bar{z}_{ij}\alpha_{ij}+\bar{\mu}_{ij}^2}{\bar{z}_{ij}}
    -\frac{\bar{z}_{ij}\alpha_{ij}\kappa}{1-\bar{z}_{ij}\alpha_{ij}+\bar{\mu}_{ij}^2}}\;,
  \end{split}
\end{equation}
the momenta after the splitting are given by
\begin{equation}
  \begin{split}
    p_i^\mu=&\;\bar{z}\;
    \frac{\tilde{p}_{ij}^\mu-\bar{\mu}_{ij}^2\tilde{K}^\mu}{
      \bar{z}_{ij}\,v_{\tilde{p}_{ij},\tilde{K}}}\,
    +\Big(y(1-\bar{z})(1+\mu_{ij}^2-\mu_i^2-\mu_j^2)
       -\bar{z}(\mu_i^2+\mu_j^2)+2\mu_i^2\Big)\,
    \frac{\tilde{K}^\mu-\bar{\kappa}\,\tilde{p}_{ij}^\mu}{
      v_{\tilde{p}_{ij},\tilde{K}}/z_{ij}}
    +k_\perp^\mu\;,\\
    p_j^\mu=&\;(1-\bar{z})\;
    \frac{\tilde{p}_{ij}^\mu-\bar{\mu}_{ij}^2\tilde{K}^\mu}{
      \bar{z}_{ij}\,v_{\tilde{p}_{ij},\tilde{K}}}\,
    +\Big(y\,\bar{z}\,(1+\mu_{ij}^2-\mu_i^2-\mu_j^2)
       -(1-\bar{z})(\mu_i^2+\mu_j^2)+2\mu_j^2\Big)\,
    \frac{\tilde{K}^\mu-\bar{\kappa}\,\tilde{p}_{ij}^\mu}{
      v_{\tilde{p}_{ij},\tilde{K}}/z_{ij}}
    -k_\perp^\mu\;.
  \end{split}
\end{equation}
The transverse momentum squared is given by
\begin{equation}
  {\rm k}_\perp^2=2\tilde{p}_{ij}\tilde{K}\,
  \Big[\,\bar{z}(1-\bar{z})\alpha_{ij}
    -(1-\bar{z})\mu_i^2-\bar{z}\mu_j^2\,\Big]\;.
\end{equation}
The construction of the transverse momentum vector proceeds as described in Sec.~\ref{sec:radiation_kinematics}.
While the choice of the reference vector defining $n_\perp^\mu$ is in principle arbitrary, it can be made
conveniently, e.g., \ to aid the implementation of spin correlations in collinear gluon splittings.

\section{Phase space factorization}
\label{sec:ps_factorization}
In this section, we discuss the factorization of the differential $n+1$ particle phase-space element
into a differential $n$ particle phase-space element and the radiative phase space. We start from
the generic four-dimensional expression for the initial-state momenta $p_a^\mu$ and $p_b^\mu$ and 
final-state momenta, $\{p_1^\mu,\ldots,p_i^\mu,\ldots,p_j^\mu,\ldots,p_n^\mu\}$, 
\begin{equation}\label{eq:final_phase_space}
  \begin{split}
    {\rm d}\Phi_{n}(p_a,p_b;p_1,\ldots,p_i,\ldots,p_j,\ldots,p_n)
    =&\;\left[\prod_{i=1}^{n}\frac{1}{(2\pi)^3}
    \frac{{\rm d}^3p_i}{2p_i^0}\right]
    (2\pi)^4\delta^{(4)}\Big(p_a+p_b-\sum p_i\Big)\;.
  \end{split}
\end{equation}
Replacing particles $i$ and $j$ with a combined ``mother'' particle $\widetilde{\imath\jmath}$,
we can write the expression for the underlying Born phase space as
\begin{equation}\label{eq:born_phase_space}
  \begin{split}
    &{\rm d}\Phi_{n-1}(\tilde{p}_a,\tilde{p}_b;\tilde{p}_1,\ldots,\tilde{p}_{ij},
      \ldots,\backslash\!\!\!\!\tilde{p}_j,\ldots,\tilde{p}_n)
    =\\&\qquad\;\Bigg[\prod_{\substack{k=1\\k\neq i,j}}^{n}\frac{1}{(2\pi)^3}
    \frac{{\rm d}^3\tilde{p}_i}{2\tilde{p}_i^0}\Bigg]\frac{1}{(2\pi)^3}
    \frac{{\rm d}^3\tilde{p}_{ij}}{2\tilde{p}_{ij}^0}
    (2\pi)^4\delta^{(4)}\Big(\tilde{p}_a+\tilde{p}_b
      -\sum_{k\neq i,j}\tilde{p}_k-\tilde{p}_{ij}\Big)\;.
  \end{split}
\end{equation}
The ratio of differential phase-space elements after and before the mapping is
defined as the differential phase-space element for the one-particle emission process
\begin{equation}\label{eq:emission_phase_space}
  {\rm d}\Phi_{+1}(\tilde{p}_a,\tilde{p}_b;\tilde{p}_1,\ldots,
  \tilde{p}_{ij},\ldots,\tilde{p}_n;p_i,p_j)=
  \frac{{\rm d}\Phi_{n}(p_a,p_b;p_1,\ldots,p_i,\ldots,p_j,\ldots,p_n)}{
    {\rm d}\Phi_{n-1}(\tilde{p}_a,\tilde{p}_b;\tilde{p}_1,\ldots,\tilde{p}_{ij},
    \ldots,\backslash\!\!\!\!\tilde{p}_{j},\ldots,\tilde{p}_n)}\;.
\end{equation}
This expression naturally generalizes to $D$ dimensions.
It can be computed using the lowest final-state multiplicity, i.e., \ $n=3$.
In order to do so, we first derive a suitable expression for the $D$-dimensional
two-particle phase space in the frame of an arbitrary, time-like momentum $n$. It reads
\begin{equation}\label{eq:two-body_ps_1}
  \begin{split}
    {\rm d}\Phi_2(P;p,q)=&\;\frac{1}{(2\pi)^{2D-2}}\frac{{\rm d}^{D-1}p}{2p^0}
    \frac{{\rm d}^{D-1}q}{2q^0}\,(2\pi)^D\delta^{D}(P-p-q)
    =\frac{1}{4(2\pi)^{D-2}}\,\frac{|\vec{p}\,|^{D-2}}{
      P^0|\vec{p}\,|-p^0|\vec{P}|\cos{\theta}_{pn}}\,{\rm d}\Omega_{p,n}^{D-2}\;.
  \end{split}
\end{equation}
where ${\rm d}\Omega_{p,n}$ is the integral over the $D-1$ dimensional solid angle of $p^\mu$
in the frame of $n^\mu$. All energies and momenta are computed in this frame, and we have defined
$P=p+q$. We can re-write the momentum-dependent denominator factor on the right-hand side of
Eq.~\eqref{eq:two-body_ps_1} as
\begin{equation}
    \frac{p^0}{|\vec{p}|}\left(P^0p^0-|\vec{p}||\vec{P}|\cos{\theta}_{pn}\right)
    -P^0|\vec{p}|\left(\frac{(p^0)^2}{|\vec{p}|^2}-1\right)=
    \frac{(pn)(Pp)-p^2(Pn)}{\sqrt{(pn)^2-p^2n^2}}\;.
\end{equation}
Inserting this into Eq.~\eqref{eq:two-body_ps_1}, and using $D=4-2\eps$,
we obtain a manifestly covariant form of the phase-space element,
\begin{equation}\label{eq:two_body_ps_ddim}
    {\rm d}{\Phi}_2(P;p,q)=\frac{(4\pi^2 n^2)^{\eps}}{16\pi^2}\,
    \frac{((pn)^2-p^2n^2)^{3/2-\eps}}{((pn)(Pp)-p^2(Pn))n^2}\,{\rm d}\Omega_{p,n}^{2-2\eps}\;.
\end{equation}

\subsection{Radiation kinematics}
\label{sec:ps_factorization_rad_fi}
To derive the emission phase space for final-state splittings in the radiation kinematics
of Sec.~\ref{sec:radiation_kinematics}, we make use of the standard factorization formula
\begin{equation}\label{eq:emission_phase_space_fi_rad}
    {\rm d}\Phi_{3}(-K;p_i,p_j,q)
    ={\rm d}\Phi_{2}(-K;p_j,-n)\,
    \frac{{\rm d}n^2}{2\pi}\,
    {\rm d}\Phi_{2}(-n;p_i,q)\;,
\end{equation}
where $q=\sum_{k\neq i,j}p_k$ is the sum of all final-state momenta except $p_i$ and $p_j$.
We can use Eq.~\eqref{eq:two_body_ps_ddim} to relate the phase-space factor
${\rm d}\Phi_2(-n;p_i,q)$ to the underlying Born phase space as follows
\begin{equation}\label{eq:born_remap_fi_rad}
  \begin{split}
    \frac{{\rm d}\Phi_2(-n;{p}_i,q)}{
    {\rm d}\Phi_2(-\tilde{K};\tilde{p}_{ij},\tilde{q})}
    =&\;\bigg(\frac{(p_i n)^2-p_i^2n^2}{
      (\tilde{p}_{ij}\tilde{K})^2-\tilde{p}_{ij}^2\tilde{K}^2}\bigg)^{3/2-\eps}\;
    \frac{(\tilde{p}_{ij}\tilde{K})^2
      -\tilde{p}_{ij}^2\tilde{K}^2}{({p}_in)^2-{p}_i^2n^2}
    =\bigg(\frac{(p_i n)^2-p_i^2n^2}{
      (\tilde{p}_{ij}\tilde{K})^2-\tilde{p}_{ij}^2\tilde{K}^2}\bigg)^{1/2-\eps}\;.
  \end{split}
\end{equation}
The differential two-body phase-space element for the production of
$n^\mu$ and $p_j^\mu$ is given by
\begin{equation}\label{eq:pj_production_ps_ff}
  \begin{split}
    {\rm d}{\Phi}_2(-K;n,p_j)=&\;\frac{(4\pi^2)^{\eps}}{16\pi^2}\,
    \frac{\big((p_jn)^2-p_j^2n^2\big)^{1/2-\eps}}{(n^2)^{1-\eps}}\,{\rm d}\Omega_{j,n}^{2-2\eps}\\
    =&\;\frac{(2\tilde{p}_{ij}\tilde{K})^{-\eps}}{(16\pi^2)^{1-\eps}}\,
    \frac{\big[(1-z+\mu_{ij}^2-\mu_i^2+\mu_j^2)v_{p_j,n}\big]^{1-2\eps}}{
      2(1-z+\kappa+\mu_{ij}^2-\mu_i^2)^{1-\eps}}\,{\rm d}\Omega_{j,n}^{2-2\eps}\;,
  \end{split}
\end{equation}
Finally, we combine Eqs.~\eqref{eq:born_remap_fi_rad} and~\eqref{eq:pj_production_ps_ff}
to obtain the single-emission phase space element in Eq.~\eqref{eq:emission_phase_space}
\begin{equation}\label{eq:emission_phase_space_if}
  \begin{split}
  &{\rm d}\Phi_{+1}(\tilde{p}_a,\tilde{p}_b;\ldots,\tilde{p}_{ij},\ldots;p_i,p_j)\\
  &\quad=\bigg(\frac{2\tilde{p}_{ij}\tilde{K}}{16\pi^2}\bigg)^{1-\eps}\,
  \bigg(\frac{v_{p_i,n}v_{p_j,n}}{v_{\tilde{p}_{ij},\tilde{K}}}\bigg)^{1-2\eps}
  \frac{\big(z(1-z+\mu_{ij}^2-\mu_i^2+\mu_j^2)\big)^{1-2\eps}}{
      (1-z+\kappa+\mu_{ij}^2-\mu_i^2)^{1-\eps}}\,
      {\rm d}z\,\frac{{\rm d}\Omega_{j,n}^{2-2\eps}}{4\pi}\;.
  \end{split}
\end{equation}
In the massless limit, this simplifies to
\begin{equation}\label{eq:emission_phase_space_if_ml}
  {\rm d}\Phi_{+1}(\tilde{p}_a,\tilde{p}_b;\ldots,\tilde{p}_{ij},\ldots;p_i,p_j)
  =\bigg(\frac{2\tilde{p}_{ij}\tilde{K}}{16\pi^2}\bigg)^{1-\eps}\,
  \frac{\big(z(1-z)\big)^{1-2\eps}}{(1-z+\kappa)^{1-\eps}}\,{\rm d}z\,
      \frac{{\rm d}\Omega_{j,n}^{2-2\eps}}{4\pi}\;.
\end{equation}
We demonstrate in App.~\ref{sec:cs_comparison} that this expression
agrees with the result derived in Ref.~\cite{Catani:1996vz}.

\subsection{Splitting kinematics}
\label{sec:ps_factorization_split_fi}
To derive the emission phase space for final-state radiation in the splitting kinematics
of Sec.~\ref{sec:splitting_kinematics}, the recoiler is chosen to be the sum of the
remaining final-state partons. We make use of the standard factorization formula
\begin{equation}
    {\rm d}\Phi_{3}(q;p_i,p_j,K)
    ={\rm d}\Phi_{2}(q;K,p_{ij})\,
    \frac{{\rm d}p_{ij}^2}{2\pi}\,
    {\rm d}\Phi_{2}(p_{ij};p_i,p_j)\;,
\end{equation}
where $q=\sum_{k\neq i,j}p_k$ is the sum of final-state momenta except $p_i$ and $p_j$.
Working in the frame of $q=\tilde{K}+\tilde{p}_{ij}$, we can use Eq.~\eqref{eq:two_body_ps_ddim}
to relate the phase-space factor ${\rm d}\Phi_2(q;K,p_{ij})$ to the underlying Born differential
phase space element ${\rm d}\Phi_2(q;\tilde{K},\tilde{p}_{ij})$ as follows
\begin{equation}\label{eq:born_remap_fi_split}
  \begin{split}
    \frac{{\rm d}\Phi_2(q;K,p_{ij})}{
    {\rm d}\Phi_2(q;\tilde{K},\tilde{p}_{ij})}
    =&\;\bigg(\frac{(Kp_{ij})^2-K^2p_{ij}^2}{
      (\tilde{K}\tilde{p} _{ij})^2-\tilde{K}^2\tilde{p}_{ij}^2}\bigg)^{1/2-\eps}
    =(1-y)^{1-2\eps}(1+\mu_{ij}^2-\mu_i^2-\mu_j^2)^{1-2\eps}
    \bigg(\frac{v_{p_{ij},K}}{v_{\tilde{p}_{ij},\tilde{K}}}\bigg)^{1-2\eps}\;,
  \end{split}
\end{equation}
The decay of the intermediate off-shell parton is associated with the
differential phase-space element
\begin{equation}\label{eq:pij_decay_fi_split}
  \begin{split}
    {\rm d}{\Phi}_2(p_{ij};p_i,p_j)=&\;\frac{(4\pi^2)^{\eps}}{16\pi^2}\,
    \frac{((p_ip_j)^2-p_i^2p_j^2)^{1/2-\eps}}{(p_{ij}^2)^{1-\eps}}\,{\rm d}\Omega_{i,ij}^{2-2\eps}\\
    =&\;\frac{(2\tilde{p}_{ij}\tilde{K})^{-\eps}}{(16\pi^2)^{1-\eps}}\,
    \frac{\big(y^2(1+\mu_{ij}^2-\mu_i^2-\mu_j^2)^2-4\mu_i^2\mu_j^2\big)^{1/2-\eps}}{
    2\big(y(1+\mu_{ij}^2-\mu_i^2-\mu_j^2)+\mu_i^2+\mu_j^2\big)^{1-\eps}}\,
    {\rm d}\Omega_{i,ij}^{2-2\eps}\;.
  \end{split}
\end{equation}
Finally, we combine Eqs.~\eqref{eq:born_remap_fi_split} and~\eqref{eq:pij_decay_fi_split}
to obtain
\begin{equation}\label{eq:emission_phase_space_fi_split}
  \begin{split}
    &{\rm d}\Phi_{+1}(\tilde{p}_a,\tilde{p}_b;\tilde{p}_1,\ldots,
      \tilde{p}_{ij},\ldots,\tilde{p}_n;p_i,p_j)\\
    &\qquad=\bigg(\frac{2\tilde{p}_{ij}\tilde{K}}{16\pi^2}\bigg)^{1-\eps}\,
    (1-y)^{1-2\eps}(1+\mu_{ij}^2-\mu_i^2-\mu_j^2)^{2-2\eps}
    \bigg(\frac{v_{p_{ij},K}}{v_{\tilde{p}_{ij},\tilde{K}}}\bigg)^{1-2\eps}\\
    &\qquad\qquad\times
    \frac{\big(y^2(1+\mu_{ij}^2-\mu_i^2-\mu_j^2)^2-4\mu_i^2\mu_j^2\big)^{1/2-\eps}}{
    \big(y(1+\mu_{ij}^2-\mu_i^2-\mu_j^2)+\mu_i^2+\mu_j^2\big)^{1-\eps}}\,
    {\rm d}y\,\frac{{\rm d}\Omega_{i,ij}^{2-2\eps}}{4\pi}\;.
  \end{split}
\end{equation}
In the massless limit, this simplifies to
\begin{equation}\label{eq:emission_phase_space_fi_split_ml}
  \begin{split}
    {\rm d}\Phi_{+1}(\tilde{p}_a,\tilde{p}_b;\tilde{p}_1,\ldots,
      \tilde{p}_{ij},\ldots,\tilde{p}_n;p_i,p_j)
    =\bigg(\frac{2\tilde{p}_{ij}\tilde{K}}{16\pi^2}\bigg)^{1-\eps}\,
    (1-y)^{1-2\eps}\,y^{-\eps}\,
    {\rm d}y\,\frac{{\rm d}\Omega_{i,ij}^{2-2\eps}}{4\pi}\;.
  \end{split}
\end{equation}
In App.~\ref{sec:cdst_comparison}, we will show the equivalence of
Eq.~\eqref{eq:emission_phase_space_fi_split} to the single-emission
differential phase-space element of~\cite{Catani:2002hc}.

\section{Infrared subtraction terms at next-to-leading order}
\label{sec:nlo_subtraction}
The matching of parton showers to fixed-order NLO calculations in dimensional regularization 
based on the MC@NLO algorithm~\cite{Frixione:2002ik} requires the knowledge of integrated splitting
functions in $D=4-2\eps$ dimensions. Since our technique for massive parton evolution is modeled 
on the Catani-Seymour identified particle subtraction and the \Alaric parton-shower,
we can use the methods developed in~\cite{Hoche:2018ouj,Liebschner:2020aa}.
We will limit the discussion to the main changes needed to implement the algorithm
for massive partons. The results of this section provide an extension of the subtraction method for
identified hadrons first introduced in~\cite{Catani:1996vz}.

\subsection{Soft angular integrals}
\label{sec:angular_integrals}
In this section, we compute the angular integrals of the partial fractioned soft eikonal.
While we focus on pure final-state radiation, the results of this integration are generic
and apply to the case of final-state and initial-state emission of massless vector bosons.
The integrand is given by Eq.~\eqref{eq:pf_angular_radiator},
\begin{equation}
\begin{split}
    \bar{W}_{ik,j}^i=&\;
    \frac{l_{ik}^2}{2(l_il_j)(l_{ik}l_j)}
        -\frac{l_i^2}{2(l_il_j)^2}
        -\frac{l_k^2}{2(l_il_j)(l_kl_j)}\\
\end{split}
\end{equation}
We can write the angular phase space for the emission of the massless particle $j$ as
\begin{equation}
   \int{\rm d}\Omega_j^{2-2\eps}=\Omega(1-2\eps)
   \int{\rm d}\cos\theta(\sin^2\theta)^{-\eps}\int{\rm d}\phi(\sin^2\phi)^{-\eps}\;,
\end{equation}
where $\Omega(n)=2\pi^{n/2}/\Gamma(n/2)$ is the $d$-dimensional line element,
in particular $\Omega(1-2\eps)=2(4\pi)^{-\eps}\Gamma(1-\eps)/\Gamma(1-2\eps)$.
We finally find the following expression for the cases with massive emitter
\begin{equation}\label{eq:angular_integral_ms_ms}
  \begin{split}
    &\int\frac{{\rm d}\Omega_j^{2-2\eps}}{\Omega(1-2\eps)}\;\bar{W}_{ik,j}^i
    =\frac{l_{ik}^2}{4}I_{1,1}^{(2)}\left(\frac{l_il_{ik}}{2},l_i^2,\frac{l_{ik}^2}{4}\right)
    -\frac{l_i^2}{2}I_{2}^{(1)}(l_i^2)-\frac{l_k^2}{2}I_{1,1}^{(2)}\left(l_il_k,l_i^2,l_k^2\right)\;.
  \end{split}
\end{equation}
For massless emitter, we obtain (see also~\cite{Catani:1996vz,Herren:2022jej})
\begin{equation}\label{eq:angular_integral_ml_ms}
  \begin{split}
    &\int\frac{{\rm d}\Omega_j^{2-2\eps}}{\Omega(1-2\eps)}\;\bar{W}_{ik,j}^i
    =\frac{l_{ik}^2}{4}I_{1,1}^{(1)}\left(\frac{l_il_{ik}}{2},\frac{l_{ik}^2}{4}\right)
    -\frac{l_k^2}{2}I_{1,1}^{(1)}(l_il_k,l_k^2)\;,
  \end{split}
\end{equation}
The angular integrals $I_{1,1}$ and $I_2$ have been computed
in~\cite{vanNeerven:1985xr,Beenakker:1988bq,Somogyi:2011ir,
  Isidori:2020acz,Lyubovitskij:2021ges}. They read~\cite{Lyubovitskij:2021ges}
\begin{equation} 
    \begin{split}
        I_{1,1}^{(2)}(v_{12},v_{11},v_{22})={}& \frac{\pi}{\sqrt{v_{12}^2-v_{11}v_{22}}}
        \left\{\log{\frac{v_{12}+\sqrt{v_{12}^2-v_{11}v_{22}}}{v_{12}-\sqrt{v_{12}^2-v_{11}v_{22}}}}
        -\eps\left(\frac{1}{2}\log^2{\frac{v_{11}}{v_{13}^2}}-\frac{1}{2}\log^2\frac{v_{22}}{v_{23}^2}\right.\right.\\
        &\left.\left.+2{\rm Li}_2\left(1-\frac{v_{13}}{1-\sqrt{1-v_{11}}}\right)
        +2{\rm Li}_2\left(1-\frac{v_{13}}{1+\sqrt{1-v_{11}}}\right)\right.\right.\\
        &\left.\left.-2{\rm Li}_2\left(1-\frac{v_{23}}{1-\sqrt{1-v_{22}}}\right)
        -2{\rm Li}_2\left(1-\frac{v_{23}}{1+\sqrt{1-v_{22}}}\right)\right)+\mathcal{O}(\eps^2)\right\}\;,\\
        I_{1,1}^{(1)}(v_{12},v_{11})={}& \frac{\pi}{v_{12}}
        \left\{-\frac{1}{\eps}-\log\frac{v_{11}}{v_{12}^2}
        -\eps\left(\frac{1}{2}\log^2\frac{v_{11}}{v_{12}^2}\right.\right.\\
        &\left.\left.+2{\rm Li}_2\left(1-\frac{v_{12}}{1-\sqrt{1-v_{11}}}\right)
        +2{\rm Li}_2\left(1-\frac{v_{12}}{1+\sqrt{1-v_{11}}}\right)\right)+\mathcal{O}(\eps^2)\right\}\;,\\
        I_{2}^{(1)}(v)={}&\frac{2\pi}{v}\left\{1+\frac{\eps}{\sqrt{1-v}}
        \log\frac{1+\sqrt{1-v}}{1-\sqrt{1-v}}+\mathcal{O}(\eps^2)\right\}\;.
    \end{split}
\end{equation}
where the velocities are defined in the notation of Ref.~\cite{Lyubovitskij:2021ges},
and where 
\begin{equation}
    \begin{split}
        v_{13}=&\;(1-\lambda) v_{11}+\lambda v_{12}\;,\qquad
        v_{23}=(1-\lambda) v_{12}+\lambda v_{22}\;,\qquad
        \lambda=\frac{v_{11}-v_{12}-\sqrt{v_{12}^2-v_{11}v_{22}}}{
          v_{11}+v_{22}-2v_{12}}\;.
    \end{split}
\end{equation}
Note that spurious collinear poles appear on the right-hand side
of Eq.~\eqref{eq:angular_integral_ml_ms}, which cancel between
$I_{1,1}^{(1)}(l_il_{ik}/2,l_{ik}^2/4)/2$ and $I_{1,1}^{(1)}(l_il_k,l_k^2)$.
In addition, in the fully massless case $I_{1,1}^{(1)}$ is related to the
massless two-denominator integral and gives the simple result~\cite{Herren:2022jej}
\begin{equation}
    \begin{split}
        I_{1,1}^{(1)}(v_{12},v_{12})=&\;
        -\frac{\pi}{v_{12}^{1+\eps}}\left\{\frac{1}{\eps}
        +\eps\,{\rm Li}_2\left(1-v_{12}\right)+\mathcal{O}(\eps^2)\right\}\;.
    \end{split}
\end{equation}

\subsection{Soft energy integrals}
\label{sec:energy_integrals}
In this section, we introduce the basic techniques to solve the energy integrals
in Eq.~\eqref{eq:emission_phase_space_if}. After performing the angular integrals,
we are left with the additional $z$-dependence induced by the energy denominator
in Eq.~\eqref{eq:massive_eikonal_split}. We focus on the cases relevant for QCD
and QED soft radiation, where $\mu_{ij}=\mu_i$ and $\mu_j=0$.
The differential emission probability per dipole is then given by
\begin{equation}\label{eq:single_soft_counterterm}
  \begin{split}
  &{\rm d}\Phi_{+1}(\tilde{p}_a,\tilde{p}_b;\ldots,\tilde{p}_{ij},\ldots;p_i,p_j)\,
  \frac{n^2}{(p_jn)^2}\,\bar{W}_{ik,j}^i\\
  &\quad=\frac{(4\pi)^{\eps}}{16\pi^2}\,
  \frac{\Gamma(1-\eps)}{\Gamma(1-2\eps)}\,(2\tilde{p}_{ij}\tilde{K})^{-\eps}\,
  \bigg(\frac{v_{p_i,n}}{v_{\tilde{p}_{ij},\tilde{K}}}\bigg)^{1-2\eps}
  \frac{z^{1-2\eps}}{(1-z+\kappa)^{-\eps}}\,
      \frac{{\rm d}z}{(1-z)^{1+2\eps}}\,\frac{2}{\pi}
  \frac{{\rm d}\Omega_{j,n}^{2-2\eps}}{\Omega(1-2\eps)}\bar{W}_{ik,j}^i\;.
  \end{split}
\end{equation}
In order to carry out the integration over the energy fraction, $z$,
we expand the integrand into a Laurent series. The differential soft 
subtraction counterterm, summed over all dipoles, is given by
\begin{equation}\label{eq:soft_subtraction_term}
  \begin{split}
  {\rm d}\sigma^S_{S}=&\;-8\pi\alpha_s\mu^{2\eps}\sum_{\widetilde{\imath\jmath},k}
  {\rm d}\Phi_{+1}(\tilde{p}_a,\tilde{p}_b;\ldots,\tilde{p}_{ij},\ldots;p_i,p_j)\,
  {\bf T}_{\widetilde{\imath\jmath}}{\bf T}_k\frac{n^2}{(p_jn)^2}\,\bar{W}_{ik,j}^i\\
  =&\;-\frac{\alpha_s}{2\pi}\frac{1}{\Gamma(1-\eps)}
  \sum_{\widetilde{\imath\jmath},k}
  \frac{{\bf T}_{\widetilde{\imath\jmath}}{\bf T}_k}{{\bf T}_{\widetilde{\imath\jmath}}^2}
  \bigg(\frac{4\pi\mu^2}{2p_ip_k}\bigg)^{\eps}
      \left\{-\frac{\delta(1-z)}{\eps}(1-z_-)^{-\eps}+\frac{2z}{\left[1-z\right]_+}
      -4\eps\,z\left[\frac{\log(1-z)}{1-z}\right]_+\right\}\,{\rm d}z\\
  &\;\qquad\times\,\mathcal{J}_{ik}(z,\mu_i^2,\kappa)\,
  \bigg(\frac{p_kn}{p_in}\bigg)^{\eps}
  \bigg(\frac{l_{ik}^2-l_i^2-l_k^2}{4}\bigg)^{\eps}\,
  \frac{C_{\widetilde{\imath\jmath}}}{\pi}\frac{\Gamma(1-\eps)^2}{\Gamma(1-2\eps)}\,
  \bigg[\frac{{\rm d}\Omega_{j,n}^{2-2\eps}}{\Omega(1-2\eps)}\,\bar{W}_{ik,j}^i\bigg]\;.
  \end{split}
\end{equation}
where $z_-=2\mu_i^2\big(\sqrt{1+(1+\kappa)/\mu_i^2}-1\big)$, and where the mass correction factor reads
\begin{equation}\label{eq:complete_soft_counterterm}
  \mathcal{J}_{ik}(z,\mu_i^2,\kappa)=
  \bigg(\frac{v_{p_i,n}}{v_{\tilde{p}_{ij},\tilde{K}}}\bigg)^{1-2\eps}\,.
\end{equation}
The massless limit is given by $\mathcal{J}_{ik}(z,0,\kappa)=1$.
All $1/\eps$ poles have been extracted, and we can (after expanding the
$\eps$-dependent prefactors) compute the final result as an integral over the
delta functions and plus distributions in $z$. In general, some of the terms 
must be computed numerically, as $n$ implicitly depends on $z$,
see Eq.~\eqref{eq:def_z_n_rad}.
In order to apply Eq.~\eqref{eq:complete_soft_counterterm} to processes
with resolved hadrons, we can make use of the formalism derived
in~\cite{Catani:1996vz,Hoche:2018ouj,Liebschner:2020aa}.
The $1/\eps$ pole proportional to $2z/[1-z]_+$ can then be combined
with the soft enhanced part of the collinear mass factorization counterterms.
The extension to initial-state radiation requires a repetition of the derivation
in Sec.~\ref{sec:ps_factorization_rad_fi} for initial-state kinematics.
We will discuss this in a future publication.

\subsection{Collinear integrals}
\label{sec:collinear_integrals}
To compute the purely collinear counterterms, we use the splitting kinematics
and make use of the results in App.~\ref{sec:cdst_comparison}.
Note that we also use the corresponding definitions of the scaled masses.
We define the purely collinear anomalous dimension in terms of the collinear
remainders in Eq.~\eqref{eq:coll_remainder}
\begin{equation}
  \bar{\gamma}_{ab}(\eps)=
  \frac{\Omega(2-2\eps)}{\Omega(3-2\eps)}
  \Big(\frac{z_+-z_-}{2}\Big)^{-1+2\eps}
  \int_{z_-}^{z_+}
  {\rm d}z\,\big((z-z_-)
  (z_+-z)\big)^{-\eps}\,C_{ab}(z,\eps)\;,
\end{equation}
where $z_\pm$ are given by Eq.~\eqref{eq:cdst_z_bounds}. This leads to
\begin{equation}
  \begin{split}
  \bar{\gamma}_{qq}(\eps)=&\;C_F(1-\eps)\left(1-\frac{z_++z_-}{2}\right)\;,\\
  \bar{\gamma}_{gg}(\eps)=&\;C_A\left(\frac{z_++z_-}{2}
  -\frac{(1-\eps/2)(z_++z_-)^2-z_+z_-}{3-2\eps}\right)\;,\\
  \bar{\gamma}_{gq}(\eps)=&\;T_R\left(1+\frac{z_+^2+z_-^2-(z_++z_-)}{1-\eps}
  -\frac{(z_+-z_-)^2}{3-2\eps}\right)\;.
  \end{split}
\end{equation}
The complete counterterm can now be obtained by Laurent series expansion,
using Eq.~\eqref{eq:cdst_ps}
\begin{equation}\label{eq:single_coll_counterterm}
  \begin{split}
  &\int{\rm d}\Phi_{+1}(q;\tilde{p}_{ij},\tilde{K};p_i,p_j)\frac{8\pi\alpha_s\mu^{2\eps}C_{ab}}{(p_i+p_j)^2-m_{ij}^2}\\
  &\qquad=\;\frac{\alpha_s}{2\pi}\,\bigg(\frac{\mu^2}{Q^2}\bigg)^\eps
    \frac{\Omega(3-2\eps)}{(4\pi)^{1-2\eps}}\,
    (1-\hat{\mu}_i^2-\hat{\mu}_j^2-\hat{\kappa})^{2-2\eps}
    \lambda(1,\hat{\mu}_{ij}^2,\hat{\kappa})^{-1/2+\eps}\\
  &\qquad\qquad\times\int{\rm d}y\,(1-y)^{1-2\eps}\,
    \frac{\big[\,y(1-\hat{\mu}_i^2-\hat{\mu}_j^2-\hat{\kappa})+\hat{\mu}_i^2+\hat{\mu}_j^2\,\big]^{-\eps}}{
    y(1-\hat{\mu}_i^2-\hat{\mu}_j^2-\hat{\kappa})+\hat{\mu}_i^2+\hat{\mu}_j^2-\hat{\mu}_{ij}^2}\,
  \big(z_+-z_-\big)^{1-2\eps}\,
  \bar{\gamma}_{ab}(\eps)\;.
  \end{split}
\end{equation}
There are three variants of Eq.~\eqref{eq:single_coll_counterterm}
relevant for QCD. The first describes the collinear splitting of a massive quark
into a quark and a gluon, and is characterized by $\hat{\mu}_i=\hat{\mu}_{ij}>0$ and $\hat{\mu}_j=0$.
We obtain
\begin{equation}\label{eq:single_coll_counterterm_qq}
  \begin{split}
  &\int{\rm d}\Phi_{+1}(q;\tilde{p}_{ij},\tilde{K};p_i,p_j)
  \frac{8\pi\alpha_s\mu^{2\eps}C_{qq}}{(p_i+p_j)^2-m_{ij}^2}
  =\frac{\alpha_s}{2\pi}\,
    \frac{(1-\hat{\mu}_i^2-\hat{\kappa})}{\sqrt{\lambda(1,\hat{\mu}_i^2,\hat{\kappa})}}\\
  &\qquad\times\int\frac{{\rm d}y}{y(1-\hat{\mu}_i^2-\hat{\kappa})+\hat{\mu}_i^2}\,
    \sqrt{\big[(1-y)(1-\hat{\mu}_i^2-\hat{\kappa})+2\hat{\kappa}\big]^2-4\hat{\kappa}}\;
  \bar{\gamma}_{qq}(0)\;,
  \end{split}
\end{equation}
Note that the result is infrared finite. This is expected because the only poles
that occur in the radiation off massive quarks have their origin in soft gluon
emission, which is fully accounted for by the eikonal integral in
Eq.~\eqref{eq:single_soft_counterterm}. This also shows that in our 
subtraction scheme, there is a hierarchy between the soft and the collinear
enhancements, as required for a proper classification of leading and
sub-leading logarithms.

The second case is the splitting of a gluon into two massive quarks,
which is characterized by $\hat{\mu}_i=\hat{\mu}_j>0$ and $\hat{\mu}_{ij}=0$.
The result is finite and reads
\begin{equation}\label{eq:single_coll_counterterm_gq}
  \begin{split}
  &\int{\rm d}\Phi_{+1}(q;\tilde{p}_{ij},\tilde{K};p_i,p_j)
  \frac{8\pi\alpha_s\mu^{2\eps}C_{gq}}{(p_i+p_j)^2}
  =\;\frac{\alpha_s}{2\pi}\,
    \bigg(1-\frac{2\hat{\mu}_i^2}{1-\hat{\kappa}}\bigg)\\
  &\qquad\times\int{\rm d}y\,
    \frac{\sqrt{y^2(1-2\hat{\mu}_i^2-\hat{\kappa})^2-4\hat{\mu}_i^4}}{
    (y(1-2\hat{\mu}_i^2-\hat{\kappa})+2\hat{\mu}_i^2)^2}
    \sqrt{\big[(1-y)(1-2\hat{\mu}_i^2-\hat{\kappa})+2\hat{\kappa}\big]^2-4\hat{\kappa}}\;
  \bar{\gamma}_{gq}(0)\;.
  \end{split}
\end{equation}
The final case describes the collinear decay of a massless parton
into two massless partons, and is characterized by $\hat{\mu}_i=\hat{\mu}_j=\hat{\mu}_{ij}=0$.
This term is collinearly divergent, and we obtain
\begin{equation}\label{eq:single_coll_counterterm_ml}
  \begin{split}
  &\int{\rm d}\Phi_{+1}(q;\tilde{p}_{ij},\tilde{K};p_i,p_j)
  \frac{8\pi\alpha_s\mu^{2\eps}C_{ab}}{(p_i+p_j)^2}
  =\frac{\alpha_s}{2\pi}\bigg(\frac{\mu^2}{Q^2}\bigg)^\eps
    \frac{(4\pi)^{\eps}}{\Gamma(1-\eps)}\\
  &\qquad\times\int{\rm d}y\,
    \left\{-\frac{\delta(y)}{\eps}\bigg(\frac{1-\sqrt{\hat{\kappa}}}{1+\sqrt{\hat{\kappa}}}\bigg)^{-\eps}+\frac{1}{[y]_+}\right\}
  \frac{\big(\big[(1-y)(1-\hat{\kappa})+2\hat{\kappa}\big]^2-4\hat{\kappa}\big)^{1/2-\eps}}{(1-\hat{\kappa})^{1-\eps}}\,
  \frac{\Gamma(1-\eps)^2}{\Gamma(2-2\eps)}\,
  \bar{\gamma}_{ab}(\eps)\;.
  \end{split}
\end{equation}
The treatment of initial-state singularities will be discussed in a future publication.

\section{Kinematical effects on sub-leading logarithmic corrections}
\label{sec:nll_proof}
We now demonstrate that our kinematics mapping satisfies the criteria for
NLL accuracy laid out in ~\cite{Dasgupta:2018nvj,Dasgupta:2020fwr}, extending
the discussion of the massless case in~\cite{Herren:2022jej}. We refer the reader
to~\cite{Herren:2022jej} for more details on the general method of the proof.
The analysis is based on the technique for general final-state resummation 
introduced in~\cite{Banfi:2004yd}. Following the notation of Sec.~2 
of~\cite{Banfi:2004yd}, the observable to be resummed is denoted by $v$,
while the hard partons and soft emission have momenta $p_1, \ldots, p_n$ and $k$,
respectively. The observable is a function of these momenta, $v=V(\{p\},\{k\})$,
and for any radiating color dipole formed by the hard momenta, $p_i$ and $p_j$,
the momentum of a single emission can be parametrized as
\begin{equation}\label{eq:sudakov_decomposition}
  k=z_{i,j} p_i+z_{j,i} p_j+k_{T,ij}\;,
  \qquad\text{where}\qquad
  k_{T,ij}^2=2p_ip_j\,z_{i,j}\,z_{j,i}\;.
\end{equation}
with rapidity $\eta_{ij}=1/2\ln(z_{i,j}/z_{j,i})$. An observable can be expressed as
\begin{equation}\label{eq:V_approx}
  V(k) = d_l\left(\frac{k_{T,l}}{Q}\right)^a e^{-b_l\eta_l}g_l(\phi_l)\;,
\end{equation}
with $k_{T,l}=k_{T,lj}$ and $\eta_l=\eta_{lj}$ for $j\nparallel l$ in the collinear limit.
A natural extension of Eq.~\eqref{eq:sudakov_decomposition} to the case of massive emitters
would be
\begin{equation}\label{eq:sudakov_decomposition_ms}
  k=(z_{i,j}-\bar{\mu}_j^2z_{j,i})\,p_i+(z_{j,i}-\bar{\mu}_i^2z_{i,j})\, p_j+k_{T,ij}\;,
  \qquad\text{where}\qquad
  k_{T,ij}^2=2p_ip_j\frac{2v_{p_i,p_j}}{1+v_{p_i,p_j}}\,z_{i,j}\,z_{j,i}\;.
\end{equation}
In the quasi-collinear limit, this Sudakov decomposition agrees with the
one given in~\cite{Catani:2000ef,Cacciari:2001cw} if the auxiliary (light-like)
vector defining the anti-collinear direction is chosen as the direction of the
color partner of the QCD dipole. In particular, for constant $z_{i,j}$, $z_{j,i}$
and small $k_T$, the gluon momentum behaves as
\begin{equation}
  k\overset{k_T^2\ll p_ip_j}{\underset{m_{i/j}^2\varpropto k_T^2}{\longrightarrow}}
  z_{i,j}p_i+z_{j,i}p_j+k_{T,ij}+\mathcal{O}(k_T^2)\;,
\end{equation}
in complete agreement with Eq.~\eqref{eq:sudakov_decomposition}. 
In the quasi-collinear limit, the value of the observable $V(k)$ will 
therefore be unchanged from the case of massless evolution. 
However, both the Sudakov radiator and the $\mathcal{F}$ function 
will change, due to the modified splitting functions,
Eq.~\eqref{eq:dglap_splittings}, and the effects of masses 
on the integration boundaries.

The Lund plane for gluon emission off a dipole containing a massive quark
with mass $m$ and energy $E$ will have a smooth upper rapidity bound
at $\eta=\ln(E/m)$, consistent with the well known dead-cone 
effect~\cite{Marchesini:1989yk,Dokshitzer:1991fd,Ghira:2023bxr}. When the similarity
transformation introduced in Sec.~2.2.3 of~\cite{Banfi:2004yd} is generalized
to the quasi-collinear limit, and applied to a process with massive quarks,
the location of this boundary in relative rapidity is unchanged, because
the quasi-collinear limit requires $m\varpropto k_T$. The sub-leading 
logarithmic terms in the resummed cross section can then be extracted 
similarly to the massless case, but a change to the CMW scheme is required.
We will not discuss the complete structure of the result, but address only
the effects related to momentum mapping, which have been found to spoil NLL
precision~\cite{Dasgupta:2018nvj,Dasgupta:2020fwr}.

In order to prove that the momentum mapping of Sec.~\ref{sec:radiation_kinematics}
satisfies the criteria laid out in~\cite{Dasgupta:2018nvj,Dasgupta:2020fwr},
we need to show that it has the same topological features as in the massless 
case~\cite{Herren:2022jej}. This amounts to showing that hard, (quasi-)collinear
emissions, i.e. highly energetic emissions in the direction of the hard partons,
do not generate Lorentz transformations that change existing momenta by more than
$\mathcal{O}((k_T^2/K^2)^{\tilde{\rho}})$, where $\tilde{\rho}$ is a positive
exponent. To show this, we analyze the behavior of
Eq.~\eqref{eq:lorentz_transformation} in the (quasi-)collinear region.
We can split $K^{\mu}$ into its components along the original recoil momentum,
$\tilde{K}^{\mu}$, the emitter, $\tilde{p}_i^{\mu}$, and the emission, ${p}_j^{\mu}$.
\begin{equation}\label{eq:lt_defining_momentum}
    K^{\mu}=\tilde{K}^{\mu}-X^{\mu},
    \qquad {\rm where} \qquad X^{\mu}=p^{\mu}_j-(\tilde{p}_{ij}-p_i)\;.
\end{equation}
For the effect of the Lorentz transformation that determines the event topology
after radiation, only the parametric form of $X^\mu$ is of interest. 
Using Eqs.~\eqref{eq:def_pi_rad} and~\eqref{eq:def_pj_rad}, the small momentum
$X^{\mu}$ for gluon radiation can be written as
\begin{equation}\label{eq:def_X_rad}
  \begin{split}
  X^\mu=&\;\bigg(\frac{1-z}{v_{\tilde{p}_{ij}\tilde{K}}\zeta}-(1-\bar{z})\bigg)\,\tilde{p}_{ij}^\mu
  -\frac{\mu_{ij}^2}{v_{\tilde{p}_{ij}\tilde{K}}}\bigg(\frac{1-z}{1+v_{\tilde{p}_{ij}\tilde{K}}}\,\tilde{K}^\mu
  +\frac{1-\bar{z}^2}{\bar{z}}\left(\tilde{K}^\mu-\bar{\kappa}\,\tilde{p}_{ij}^\mu\right)\bigg)\\
  &\;+\bar{v}\,\frac{1+v_{\tilde{p}_{ij}\tilde{K}}}{2v_{\tilde{p}_{ij}\tilde{K}}}\left[\left(\tilde{K}^\mu-\bar{\kappa}\tilde{p}_{ij}^\mu\right)
  -\frac{1-\bar{z}+\bar{\kappa}}{\zeta}\left(\tilde{p}_{ij}^\mu-\bar{\mu}_{ij}^2\tilde{K}^\mu\right)\right]
  +k_\perp^\mu\;.
  \end{split}
\end{equation}
In the quasi-collinear limit, this scales as $\mathcal{O}(k_\perp)$, 
which is sufficient to make the effect of the Lorentz transformation
vanish even for hard quasi-collinear splittings.

Finally, we note that the precise treatment of transverse recoil in the 
kinematics mapping of Sec.~\ref{sec:splitting_kinematics} does not affect
the resummed result at NLL precision, because the mapping is applied solely
to the purely collinear parts of the splitting functions, Eq.~\eqref{eq:coll_remainder}.
This can be understood with the help of Eq.~(2.46) in Ref.~\cite{Banfi:2004yd},
which will be structurally similar in the case of massive partons.
The sub-leading logarithmic terms, which lead to a violation of NLL accuracy
in many dipole showers, arise from accidental correlations between multiple
soft-enhanced emissions. This means in particular, that the parton-shower
equivalent of the integrals in Eq.~(2.46) does not factorize in the
strongly ordered limit. The differential probability for the emissions
is given by the derivative of the radiator function in Eq.~(2.21)
of Ref.~\cite{Banfi:2004yd} and does not receive a contribution from the
purely collinear remainders in Eq.~\eqref{eq:coll_remainder}, such that
a change in the kinematics mapping cannot generate this particular type
of NLL violation.

\section{Comparison to experimental data}
\label{sec:comparison}
\begin{figure}[p]
  \centering
  \includegraphics[width=5.5cm]{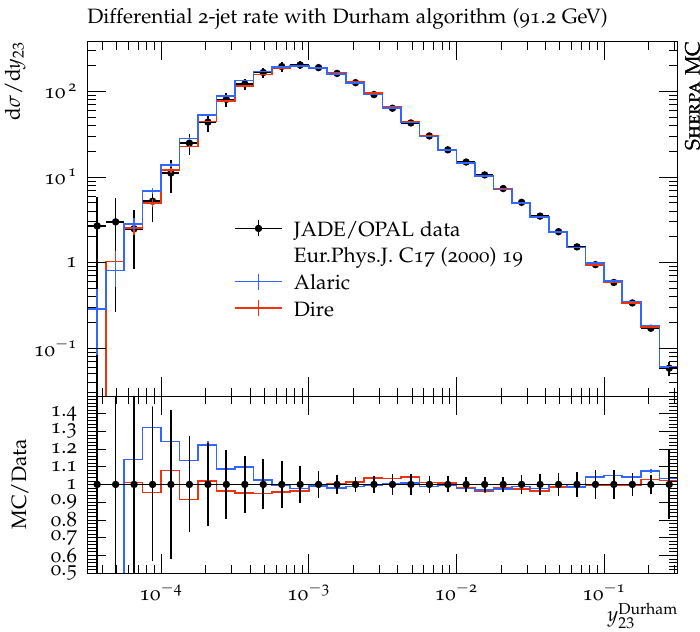}\hskip 5mm
  \includegraphics[width=5.5cm]{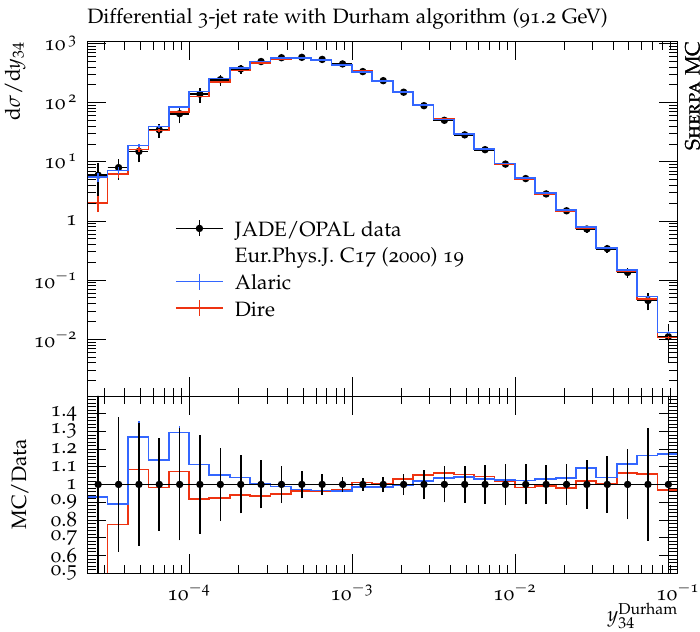}\\
  \includegraphics[width=5.5cm]{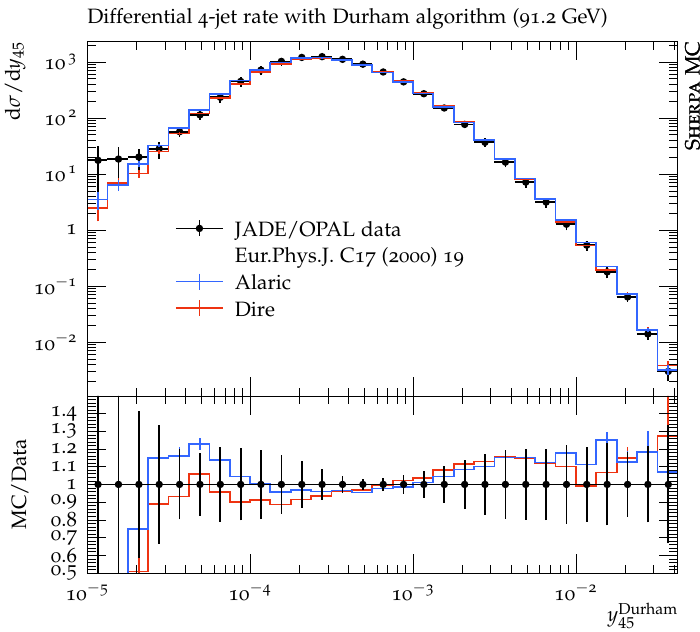}\hskip 5mm
  \includegraphics[width=5.5cm]{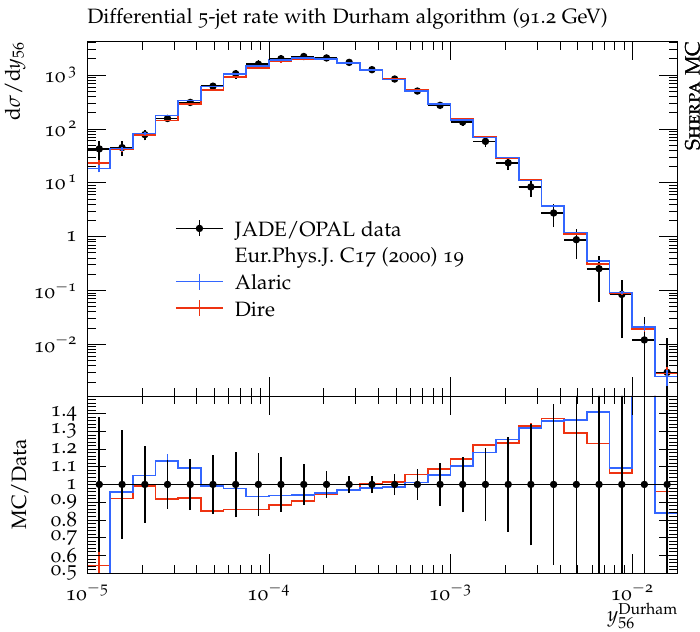}
  \caption{\Alaric and \Dire predictions in comparison to LEP data from~\cite{Pfeifenschneider:1999rz}.
    \label{fig:lep_jetrates}}
\end{figure}
\begin{figure}[p]
  \centering
  \includegraphics[width=5.5cm]{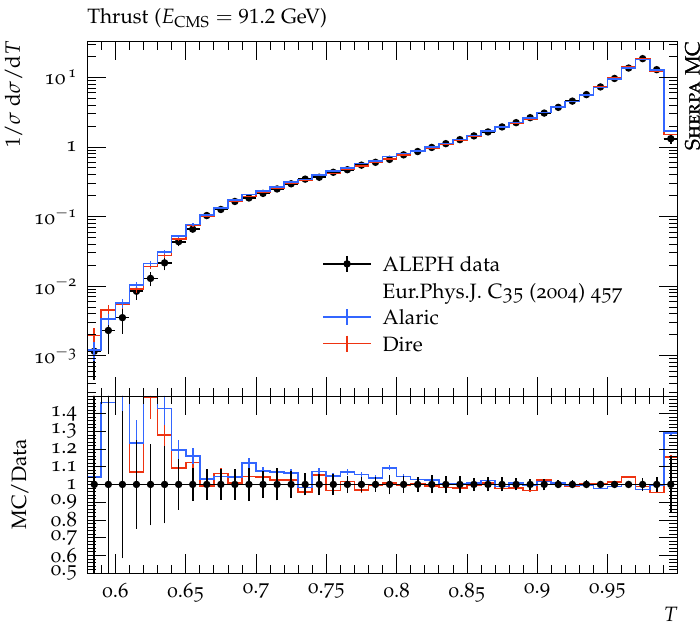}\hskip 5mm
  \includegraphics[width=5.5cm]{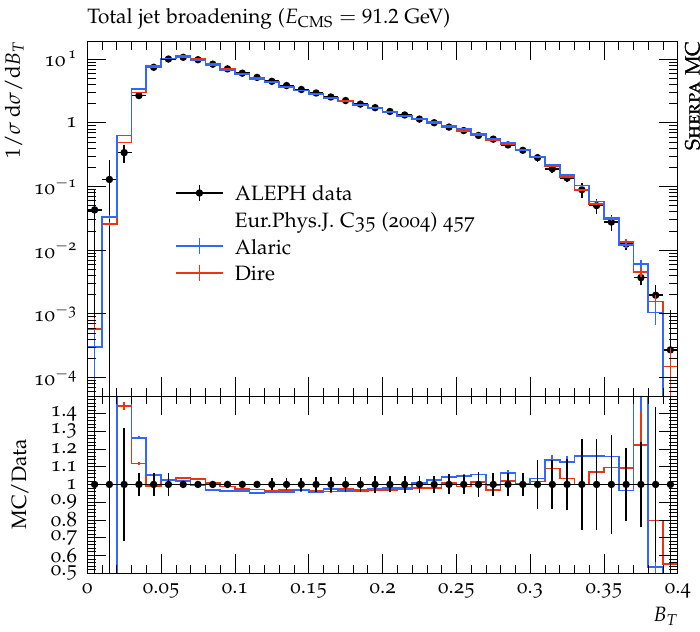}\\
  \includegraphics[width=5.5cm]{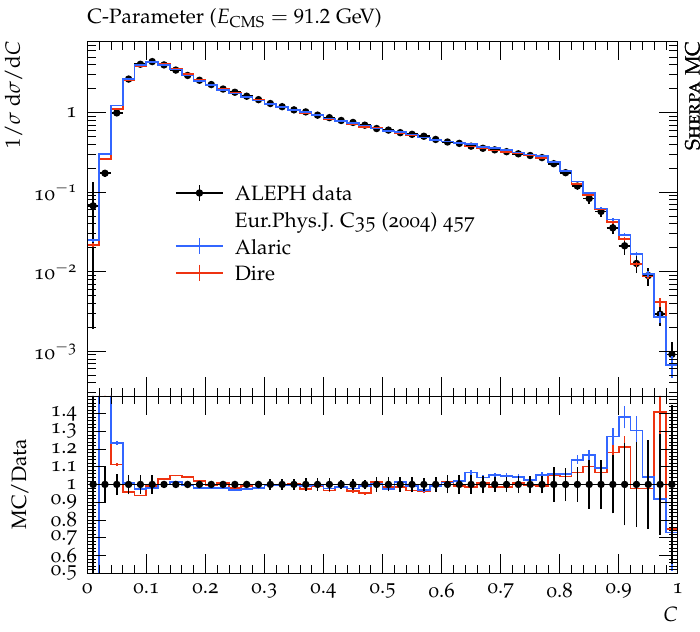}\hskip 5mm
  \includegraphics[width=5.5cm]{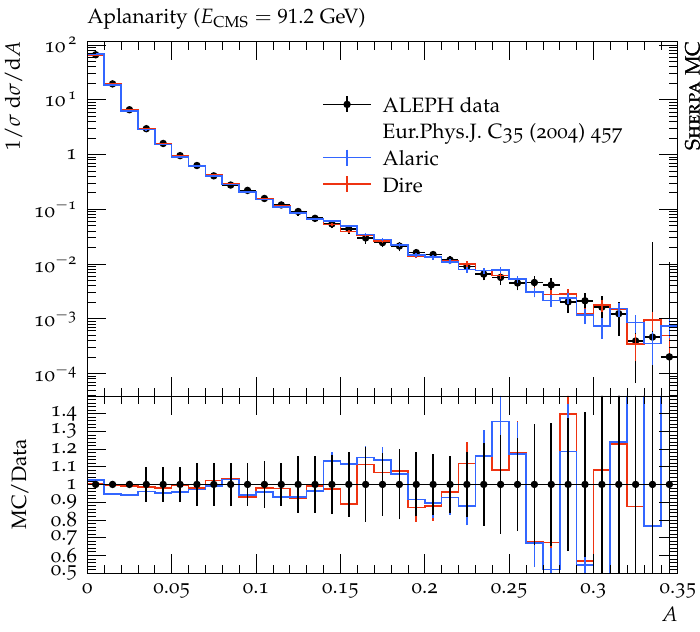}
  \caption{\Alaric and \Dire predictions in comparison to LEP data from~\cite{Heister:2003aj}.
    \label{fig:lep_shapes}}
\end{figure}
\begin{figure}[t]
  \centering
  \includegraphics[width=5.5cm]{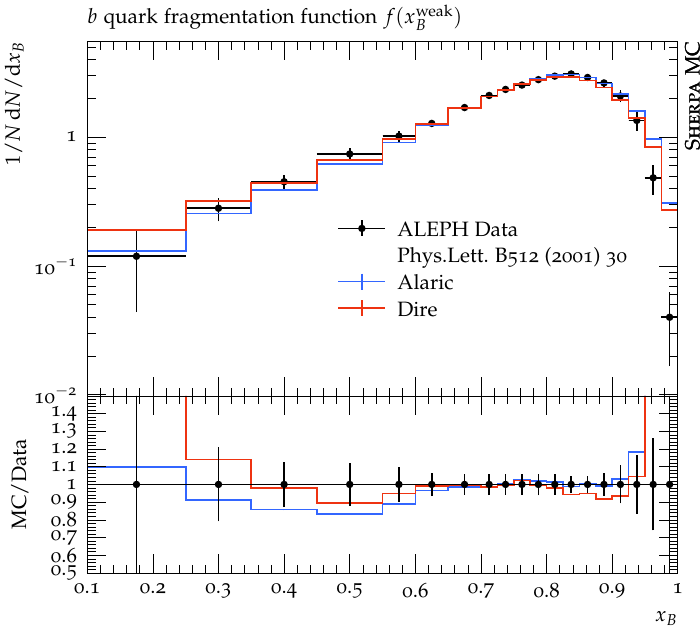}\hskip 5mm
  \includegraphics[width=5.5cm]{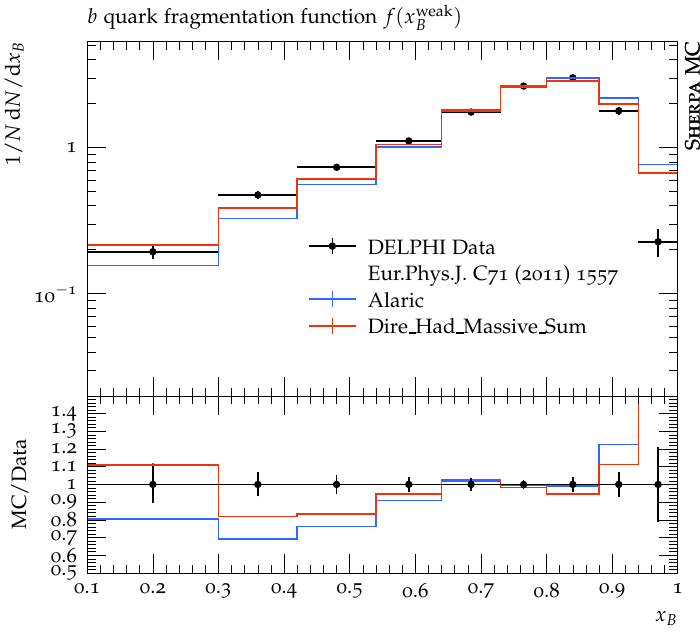}\\
  \includegraphics[width=5.5cm]{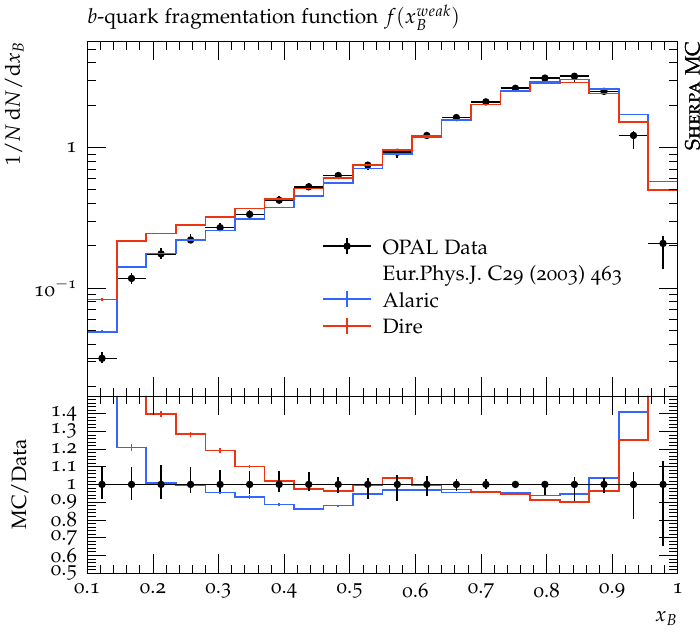}\hskip 5mm
  \includegraphics[width=5.5cm]{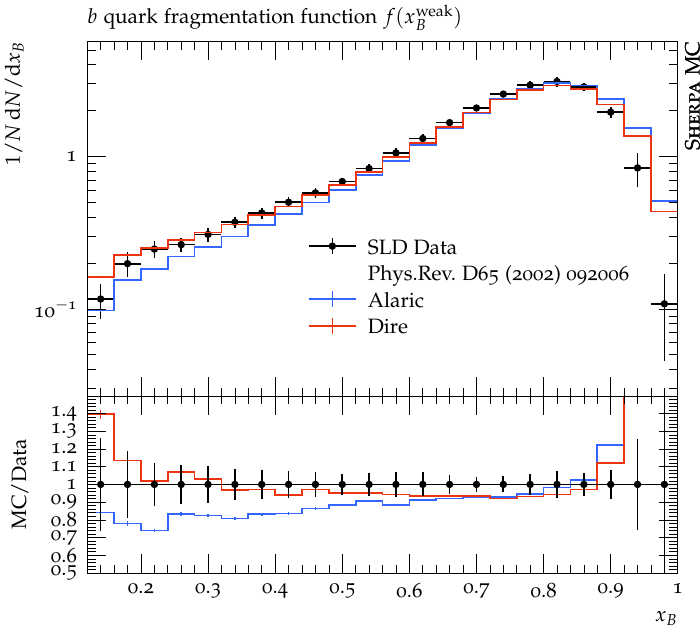}
  \caption{\Alaric and \Dire predictions in comparison to LEP data
    from~\cite{ALEPH:2001pfo,DELPHI:2011aa,OPAL:2002plk,SLD:2002poq}.
    \label{fig:lep_sld_bfrag}}
\end{figure}
In this section we present first numerical results obtained with the 
extension of the \Alaric final-state parton shower to massive parton
evolution. The implementation is part of the event generation framework
\Sherpa~\cite{Gleisberg:2003xi,Gleisberg:2008ta,Sherpa:2019gpd}. 
We do not perform NLO matching or multi-jet merging, and we set
$C_F=(N_c^2-1)/(2N_c)=4/3$ and $C_A=3$. The quark masses are set to
$m_c=1.42$~GeV and $m_b=4.75$~GeV and the same flavor thresholds
are used for the evaluation of the strong coupling. The running coupling
is evaluated at two loop accuracy, and we set $\alpha_s(m_z)=0.118$.
Following standard practice to improve the logarithmic accuracy of the 
parton shower, we employ the CMW scheme~\cite{Catani:1990rr}. In this scheme,
the soft eikonal contribution to the flavor conserving splitting functions
is rescaled by $1+\alpha_s(t)/(2\pi) K$, where $K=(67/18-\pi^2/6)\,C_A-10/9\,T_R\,n_f$,
and where $n_f$ is scale dependent with the same flavor thresholds as listed above.
Velocity dependent corrections to $K$ should be included in principle,
but the massless result provides an acceptable approximation at very large 
velocities~\cite{Korchemsky:1987wg,Kidonakis:2009ev}, and can therefore
be used for $b$-quark production at LEP, where $v\approx 0.99$.
Our results include the simulation of hadronization using the Lund string 
fragmentation implemented in Pythia~6.4~\cite{Sjostrand:2006za}.
We use the default hadronization parameters, apart from the following:
\texttt{PARJ(21)=0.3}, \texttt{PARJ(41)=0.4}, \texttt{PARJ(42)=0.45},
\texttt{PARJ(46)=0.5} and compare our predictions with those from
the \Dire parton shower~\cite{Hoche:2015sya}.
All analyses are performed with Rivet~\cite{Buckley:2010ar}.

Figure~\ref{fig:lep_jetrates} displays predictions from the \Alaric parton shower
for differential jet rates in the Durham scheme compared to experimental results
from the JADE and OPAL collaborations~\cite{Pfeifenschneider:1999rz}. 
The $b$-quark mass corresponds to $y\sim 2.8\cdot10^{-3}$, and hadronization effects
dominate the predictions below $\sim10^{-4}$. We observe good agreement with the
experimental data. Overall, the quality of the results is similar to 
Ref.~\cite{Herren:2022jej}, where heavy quark effects were modeled by
thresholds.
Figure~\ref{fig:lep_shapes} shows a comparison for event shapes measured by the
ALEPH collaboration~\cite{Heister:2003aj}. It can be expected that the numerical
predictions will improve upon including matrix-element corrections, or when merging
the parton shower with higher-multiplicity calculations. Again, the overall quality
of the prediction is similar to Ref.~\cite{Herren:2022jej}.

Finally, we show predictions for the $b$-quark fragmentation function
as measured by the ALEPH~\cite{ALEPH:2001pfo}, DELPHI~\cite{DELPHI:2011aa},
OPAL~\cite{OPAL:2002plk} and SLD~\cite{SLD:2002poq} collaborations.
Figure~\ref{fig:lep_sld_bfrag} shows a fair agreement of both the
\Alaric and \Dire predictions with experimental data. We note that
both parton shower implementations use the same hadronization tune.

\section{Conclusions}
\label{sec:conclusions}
We have introduced an extension of the recently proposed \Alaric 
parton-shower model to the case of massive QCD evolution. An essential 
aspect of the new algorithm is the form of the collinearly matched massive
eikonal, which is obtained by partial fractioning the angular component
of the eikonal of a complete dipole. This technique preserves the positivity
of the splitting function, thus leading to an excellent efficiency of the
Monte-Carlo simulation. Inspired by the symmetry of the partonic final state
in purely collinear splittings, we also introduced a dedicated kinematics 
mapping for this scenario and showed that it preserves the NLL precision
of the overall simulation. We computed the infrared counterterms needed
for the matching to fixed-order calculations at NLO accuracy, and discussed
the logarithmic structure of the resummation in the case of heavy-quark evolution. 

Several improvements of this algorithm are needed before it can be
considered on par with the parton shower simulations used by past 
and current experiments. Clearly, spin correlations and dominant
sub-leading color effects should be included. This can be achieved
with the help of the techniques from~\cite{
  Collins:1987cp,Knowles:1987cu,Knowles:1988vs,Knowles:1988hu,
  Gustafson:1992uh,Dulat:2018vuy,Hamilton:2020rcu}. 
An extension to initial-state evolution is needed for LHC phenomenology.
It will need to account for the non-cancellation of certain types of
singularities in processes with two massive initial states~\cite{Caola:2020xup}.
Finally, the algorithm should be extended to higher orders based on the
techniques developed in~\cite{Hoche:2017iem,Dulat:2018vuy,Gellersen:2021eci}. 

In this context, we note that the all-orders (in $\eps$) expressions from
Sec.~\ref{sec:nlo_subtraction}, in conjunction with higher-order expressions
for the angular integrals in Sec.~\ref{sec:angular_integrals} that can be
obtained from~\cite{Lyubovitskij:2021ges}, can be used to compute the
factorizable integrals at NNLO, thus providing a significant part of the
components needed for an MC@NNLO matching~\cite{Campbell:2021svd}.
The computation of the remaining non-factorizable integrals is a further
development needed in order to reach the precision targets of the
high-luminosity LHC.

\section*{Acknowledgments}
We thank John Campbell, Florian Herren and Simone Marzani for many helpful
and inspiring discussions. We are grateful to Davide Napoletano and Pavel Nadolsky
for clarifying various questions on PDFs and the implementation of the FONLL and ACOT schemes. 
This work was supported by the Fermi National Accelerator Laboratory (Fermilab),
a U.S. Department of Energy, Office of Science, HEP User Facility.
Fermilab is managed by Fermi Research Alliance, LLC (FRA),
acting under Contract No. DE--AC02--07CH11359.

\appendix
\section{Phase-space factorization in comparison to other infrared subtraction schemes}
In this appendix, we compare the phase-space factorization in our method to the existing
dipole subtraction schemes of Refs.~\cite{Catani:1996vz} and~\cite{Catani:2002hc}.
We will show that our generic form of the factorized phase space, derived from
Eq.~\eqref{eq:two_body_ps_ddim} can be used to obtain the relevant formulae, 
for pure final-state evolution.

\subsection{Massless radiation kinematics}
\label{sec:cs_comparison}
First, we show that Eq.~(5.189) of Ref.~\cite{Catani:1996vz} can be derived from
our generic expression, Eq.~\eqref{eq:two_body_ps_ddim}, using radiation kinematics.
We start with the massless limit of Eq.~\eqref{eq:emission_phase_space_if}
\begin{equation}
  {\rm d}\Phi_{+1}(\tilde{p}_a,\tilde{p}_b;\ldots,\tilde{p}_{ij},\ldots;p_i,p_j)
  =\bigg(\frac{2\tilde{p}_{ij}\tilde{K}}{16\pi^2}\bigg)^{1-\eps}\,
  \frac{\big(z(1-z)\big)^{1-2\eps}}{(1-z+\kappa)^{1-\eps}}\,{\rm d}z\,
      \frac{{\rm d}\Omega_{j,n}^{1-2\eps}}{4\pi}\;.
\end{equation}
Expressing the polar angle in the frame of $n$ in terms of $v$ and $z$
(see also Eq.~(32) of~\cite{Herren:2022jej})
\begin{equation}
    \cos\theta_{j,n}=1-2v\frac{1-z+\kappa}{1-z}=1-\frac{n^2\,v\,z}{(1-z)\,p_an}\;.
\end{equation}
we can perform a change of integration  variables ${\rm d}\cos\theta_{j,n}\to{\rm d}v$,
leading to Eq.~(5.189) of Ref.~\cite{Catani:1996vz}.
\begin{equation}
  \begin{split}
  {\rm d}\Phi_{+1}(\tilde{p}_a,\tilde{p}_b;\ldots,\tilde{p}_{ij},\ldots;p_i,p_j)
  =&\;\frac{(2\tilde{p}_{ij}\tilde{K})^{1-\eps}}{16\pi^2}\,
  \frac{\big(z(1-z)\big)^{1-2\eps}}{(1-z+\kappa)^{-\eps}}\,
  \frac{(\sin^2\theta_{j,n})^{-\eps}}{1-z}\,
      {\rm d}z\,{\rm d}v\,\frac{{\rm d}\Omega_{j,n}^{1-2\eps}}{(4\pi)^{1-2\eps}}\\
  =&\;\frac{(2p_in)^{1-\eps}}{16\pi^2}\,(1-z)^{-2\eps}\,
      \bigg(\frac{vz}{1-z}\bigg(1-\frac{n^2vz}{(1-z)\,2p_in}\bigg)\bigg)^{-\eps}
      {\rm d}z\,{\rm d}v\,\frac{{\rm d}\Omega_{j,n}^{1-2\eps}}{(2\pi)^{1-2\eps}}\;.
  \end{split}
\end{equation}

\subsection{Massive splitting kinematics}
\label{sec:cdst_comparison}
In this appendix, we derive the phase-space factorization formula, Eq.~(5.11)
in~\cite{Catani:2002hc} from our generic expression, Eq.~\eqref{eq:two_body_ps_ddim}.
We use the definitions of Eq.~\eqref{eq:def_y_z_cdst} \cite{Catani:2002hc}, and we set
$Q^2=(\tilde{p}_{ij}+\tilde{K})^2$. In addition, we define the scaled masses\footnote{
Note that these definitions differ from the ones in Sec.~\ref{sec:splitting_kinematics}.}
\begin{equation}
  \hat{\mu}_i^2=\frac{m_i^2}{Q^2}\;,\qquad
  \hat{\mu}_j^2=\frac{m_j^2}{Q^2}\;,\qquad
  \hat{\mu}_{ij}^2=\frac{m_{ij}^2}{Q^2}\;,\qquad
  \hat{\kappa}=\frac{K^2}{Q^2}\;.
\end{equation}
The single-emission phase space element is given by
\begin{equation}
  \begin{split}
    {\rm d}\Phi_{+1}(q;\tilde{p}_{ij},\tilde{K};p_i,p_j)=
    \frac{{\rm d}\Phi_2(q;p_{ij},K)}{{\rm d}\Phi_2(q;\tilde{p}_{ij},\tilde{K})}
    \frac{{\rm d}p_{ij}^2}{2\pi}\,{\rm d}\Phi_2(p_{ij};p_i,p_j)\;.
  \end{split}
\end{equation}
The decay of $p_{ij}$ is simplest to compute in its rest frame.
In this frame, we can write
\begin{equation}
    z=\frac{E_i^{(ij)}E_{K}^{(ij)}}{p_{ij}K}
    \Big(1-v_{ij,i}v_{ij,k}\cos\theta_{i,ij}\Big)
    =\frac{p_ip_{ij}}{p_{ij}^2}
    \Big(1-v_{ij,i}v_{ij,k}\cos\theta_{i,ij}\Big)\;,
\end{equation}
where the velocities are given by
\begin{equation}
    v_{ij,i}=\frac{\sqrt{y^2(1-\hat{\mu}_i^2-\hat{\mu}_j^2-\hat{\kappa})^2-4\hat{\mu}_i^2\hat{\mu}_j^2}}{
    y(1-\hat{\mu}_i^2-\hat{\mu}_j^2-\hat{\kappa})+2\hat{\mu}_i^2}\;,
    \qquad
    v_{ij,k}=\frac{\sqrt{\big[(1-y)(1-\hat{\mu}_i^2-\hat{\mu}_j^2-\hat{\kappa})+2\hat{\kappa}\big]^2-4\hat{\kappa}}}{
    (1-y)(1-\hat{\mu}_i^2-\hat{\mu}_j^2-\hat{\kappa})}
\end{equation}
The decay phase space, written in the frame of $p_{ij}$, then reads
\begin{equation}
  \begin{split}
    {\rm d}{\Phi}_2(p_{ij};p_i,p_j)=&\;\frac{(4\pi^2)^{\eps}}{16\pi^2}\,
    \frac{(E_i^{(ij)}v_{ij,i})^{1-2\eps}}{(p_{ij}^2)^{1/2}}\,{\rm d}\Omega_{i,ij}^{2-2\eps}
    =\frac{(4\pi^2)^{\eps}}{16\pi^2}\,
    \frac{\big(p_{i,\perp}^{(ij)}\big)^{-2\eps}}{v_{ij,k}}\,
    {\rm d}z\,{\rm d}\Omega_{i,ij}^{1-2\eps}\;.
  \end{split}
\end{equation}
We can use the factorization Ansatz
$p_{i,\perp}^{(ij)\,2}=X(z-z_-)(z_+-z)$,
where the physical boundary condition gives the roots of the quadratic
$|\cos\theta_{i,ij}|=1$, leading to
\begin{equation}\label{eq:cdst_z_bounds}
    z_\pm=\frac{p_ip_{ij}}{p_{ij}^2}
    \Big(1\pm v_{ij,i}v_{ij,k}\Big)
    =\frac{y(1-\hat{\mu}_i^2-\hat{\mu}_j^2-\hat{\kappa})+2\hat{\mu}_i^2}{
    2\big[y(1-\hat{\mu}_i^2-\hat{\mu}_j^2-\hat{\kappa})+\hat{\mu}_i^2+\hat{\mu}_j^2\big]}
    \Big(1\pm v_{ij,i}v_{ij,k}\Big)\;.
\end{equation}
The factor $X$ is determined by equating the transverse momentum at
the extremal point $z_{i,\rm max}=p_ip_{ij}/p_{ij}^2$
to the total three-momentum of $p_i$ in the frame of $p_{ij}$. This leads to
\begin{equation}
    p_{i,\perp}^{(ij)\,2}=\frac{p_{ij}^2}{v_{ij,k}^2}
    (z-z_-)(z_+-z)\;,
\end{equation}
and finally
\begin{equation}\label{eq:cdst_decay}
  \begin{split}
    {\rm d}{\Phi}_2(p_{ij};p_i,p_j)=&\;\frac{(4\pi^2/Q^2)^{\eps}}{16\pi^2\,v_{ij,k}^{1-2\eps}}\,
    [y(1-\hat{\mu}_i^2-\hat{\mu}_j^2-\hat{\kappa})+\hat{\mu}_i^2+\hat{\mu}_j^2]^{-\eps}
    \big((z-z_-)(z_+-z)\big)^{-\eps}\,
    {\rm d}z\,{\rm d}\Omega_{i,ij}^{1-2\eps}\;.
  \end{split}
\end{equation}
We also have
\begin{equation}\label{eq:cdst_prop}
    {\rm d}p_{ij}^2=Q^2(1-\hat{\mu}_i^2-\hat{\mu}_j^2-\hat{\kappa})\,{\rm d}y\;.
\end{equation}
The last missing component of the emission phase space is the ratio of the
production to the underlying Born phase-space element. It is most easily computed
in the frame of $q=\tilde{p}_{ij}+\tilde{K}$ and results in
\begin{equation}\label{eq:cdst_prod}
    \frac{{\rm d}\Phi_2(q;p_{ij},K)}{{\rm d}\Phi_2(q;\tilde{p}_{ij},\tilde{K})}
    =\bigg(\frac{(2p_{ij}K)^2\,v_{ij,k}^2}{Q^4\lambda(1,\hat{\mu}_{ij}^2,\hat{\kappa})}\bigg)^{1/2-\eps}
    =\bigg(\frac{(1-y)^2(1-\hat{\mu}_i^2-\hat{\mu}_j^2-\hat{\kappa})^2\,v_{ij,k}^2}{
    \lambda(1,\hat{\mu}_{ij}^2,\hat{\kappa})}\bigg)^{1/2-\eps}\;.
\end{equation}
where the K{\"a}llen function is given by $\lambda(a,b,c)=(a-b-c)^2-4bc$.
Combining Eqs.~\eqref{eq:cdst_decay}-\eqref{eq:cdst_prod} gives the final result
\begin{equation}\label{eq:cdst_ps}
  \begin{split}
  {\rm d}\Phi_{+1}(q;\tilde{p}_{ij},\tilde{K};p_i,p_j)
  =&\;\frac{(Q^2)^{1-\eps}}{16\pi^2}
    (1-\hat{\mu}_i^2-\hat{\mu}_j^2-\hat{\kappa})^{2-2\eps}
    \lambda(1,\hat{\mu}_{ij}^2,\hat{\kappa})^{-1/2+\eps}\\
  &\qquad\times{\rm d}y\,(1-y)^{1-2\eps}\,
    \big[\,y(1-\hat{\mu}_i^2-\hat{\mu}_j^2-\hat{\kappa})+\hat{\mu}_i^2+\hat{\mu}_j^2\,\big]^{-\eps}\\
  &\qquad\times{\rm d}z\,\big((z-z_-)
  (z_+-z)\big)^{-\eps}\;
    \frac{{\rm d}\Omega_{i,ij}^{1-2\eps}}{(2\pi)^{1-2\eps}}\;.
  \end{split}
\end{equation}

\bibliography{main}

\end{document}